\documentclass[9pt,twocolumn,twoside]{opticajnl}
\journal{opticajournal} 
\bibpunct{}{}{,}{s}{}{}
\usepackage[utf8]{inputenc}
\usepackage{amsmath}
\DeclareUnicodeCharacter{03BB}{$\lambda$}
\DeclareUnicodeCharacter{03C0}{$\pi$}
\setboolean{shortarticle}{false}

\usepackage{subcaption}
\makeatletter
\renewcommand*\subcaption@label{\caption@withoptargs\subcaption@@label}
\makeatother
\usepackage{stfloats}
\usepackage{multirow}
\usepackage{siunitx}
\usepackage{physics}
\usepackage{tikz}
\usetikzlibrary{arrows.meta,positioning,fit,backgrounds}
\usepackage{graphicx}
\usepackage{subcaption}
\DeclareSIUnit\baud{Baud}
\usepackage{amsthm} 
\usepackage{xcolor}
\definecolor{revisiongreen}{RGB}{0,128,0}
   
\usepackage{soul}

\usepackage{lineno}

\usepackage{booktabs}
\usepackage{tabularx}
\usepackage{array}
\usepackage{colortbl}

\title{Multimode continuous-variable quantum key distribution with CMOS-compatible photonic–electronic integration}

\author[1,*]{Yiming Bian}
\author[2,*]{Sarah Bastiaens}
\author[1,*]{Saeed Hakimi}
\author[1,*]{Huy Quang Nguyen}
\author[2,*]{Aboobackkar Sidhique}
\author[2]{C\'{e}dric Bruynsteen}
\author[2]{Govind J. Wanjalkar}
\author[2]{Ali Al Hadi Amhaz}
\author[2]{Jasper Jans}
\author[2]{Axl Bomhals}
\author[1]{Ulrik L. Andersen}
\author[1]{Tobias Gehring}
\author[2, $^\ddagger$]{Xin Yin}
\author[1, $\dagger$]{Adnan A.E. Hajomer}

\affil[1]{Center for Macroscopic Quantum States (bigQ), Department of Physics, Technical University of Denmark, 2800 Kongens Lyngby, Denmark}

\affil[2]{Ghent University-imec, IDLab, Dep. INTEC, 9052 Ghent, Belgium}

\affil[ ]{Corresponding authors: $^\dagger$ aaeha@dtu.dk, $^\ddagger$ xin.yin@ugent.be}

\affil[*]{These authors contributed equally as first authors}

\begin{abstract}

Multiplexing and photonic--electronic integration offer a route to scalable and high-throughput quantum key distribution (QKD) networks through parallel transmission and compact hardware suited to volume manufacturing. Realizing this route requires multichannel transmitters and receivers, and a security analysis accounting for mode coupling. Here we present a CMOS-compatible photonic--electronic multimode continuous-variable QKD architecture, using polarization multiplexing as a two-mode example. Silicon-photonic transmitter and receiver chips prepare and detect two polarization modes in parallel. The receiver is co-integrated with a dual-channel 28-nm CMOS transimpedance amplifier, providing shot-noise clearances exceeding 20~dB in both channels. Pilot-assisted unitary digital recovery removes the need for active optical polarization control, while a multimode security analysis accounts for the mode crosstalk. At 2\% residual crosstalk, the multimode analysis extends the simulated secure distance from 39 to 84~km, near the crosstalk-free limit of 89~km. Over 10~km of deployed fibre, the mean total finite-size secret-key rate estimate is 6.69~Mbit/s, 12\% above the single-mode analysis from the same digitally recovered data. These results establish a route to scalable multimode quantum-secured networks built on technology suited to mass production.

\end{abstract}
\doi{}

\setboolean{displaycopyright}{false} 

\begin{document}
  
\maketitle

\section*{Introduction}

Quantum key distribution (QKD) uses the laws of quantum mechanics to establish information-theoretically secure keys between legitimate parties \cite{gisin2002quantum,gisin2007quantum}. Continuous-variable QKD (CV-QKD) encodes information in optical field quadratures and uses coherent detection, making it compatible with optical telecommunication technologies \cite{ralph1999continuous,cerf2001quantum,grosshans2002continuous,silberhorn2002continuous}. Demonstrations of transmission from metropolitan distances to beyond 100~km \cite{jouguet2013experimental,zhang2020long,jain2022practical,hajomer2024long} and coexistence with classical data traffic \cite{eriksson2019wavelength, hajomer2025coexistence} have established important capabilities for operation within existing fibre networks. Translating these capabilities into large-scale networks requires progress in two complementary directions: compact hardware built on technology suited to volume manufacturing, and multimode transmission architectures that increase aggregate secret-key throughput and accommodate multiple users \cite{pirandola2020advances,usenko2026continuous}.

Photonic integration provides a route to compact QKD hardware by bringing optical state preparation and detection onto chips \cite{luo2023recent,wang2020integrated}. Silicon-photonic CV-QKD transmitters and receivers have been demonstrated \cite{zhang2019integrated,bian2024continuous,pietri2024experimental,ng2025gigabit,cao2026quantum,xu2026bidirectional,xu2026fully}. However, optical integration alone does not ensure the sensitive detection of weak quantum signals, which also requires low-noise electronic amplifiers with sufficient bandwidth. Several implementations connect the photonic chip to board-level amplifiers, whose noise and electrical interconnections limit not only detection performance but also system scalability \cite{ng2025gigabit,pietri2024experimental}. Integrating photonics and electronics has enabled high-speed CV-QKD using custom GaAs amplifiers \cite{hajomer2024continuous,hajomer2025chip}. Shot-noise-limited homodyne detection has also been demonstrated on a single chip using a fabrication process that supplements complementary metal–oxide–semiconductor (CMOS) electronics with bipolar transistors \cite{tasker2024bicmos}. Scaling to many parallel channels additionally requires compact electronic readout suited to volume manufacturing. Standard CMOS technology offers high circuit integration density and an established manufacturing platform. Its potential for photonic–electronic integration has been demonstrated in high-speed classical optical transceivers \cite{atabaki2018integrating}, although that work did not demonstrate the shot-noise-limited coherent detection required for CV-QKD. The challenge is to co-integrate multichannel readout implemented in a standard CMOS process with silicon photonics, without compromising the low noise and bandwidth needed for parallel quantum-signal detection.

Multiplexing complements hardware miniaturization by increasing the optical resources available for parallel key generation. Independently modulated signals can occupy orthogonal polarization, wavelength, or spatial modes, providing parallel channels for higher aggregate key rate or additional users. Wavelength-division multiplexing has enabled transmission of 194 CV-QKD channels in a single fibre \cite{eriksson2020wavelength}, while polarization-multiplexed implementations exploit the two orthogonal polarization modes of a fibre to increase the total secret-key rate \cite{roumestan2024,ng2025gigabit,fatkhiev2026monolithic}. Optical-frequency-comb networks extend wavelength multiplexing to multi-user operation, with a demonstrated network capacity of 19 users \cite{xu2026polychromatic}. These demonstrations increase the number of parallel optical channels, but their detection relies on bulk optical components or photonic circuits interfaced with separate electronic amplifiers. Extending such implementations to more modes increases hardware complexity and footprint, motivating multichannel photonic--electronic integration.

Combining multiplexing with photonic--electronic integration also requires addressing mode coupling. In dual-polarization (DP) CV-QKD, for example, polarization mixing during transmission leaves each received mode correlated with both independently modulated sources. Optical compensation can complicate integration because external polarization controllers add components, while on-chip polarization control can introduce additional optical loss \cite{zhang2019integrated,bian2024continuous,pietri2024experimental,hajomer2025chip,wang2022silicon}. Polarization-diverse coherent detection addresses this by enabling digital recovery without active optical polarization control \cite{eriksson2020wavelength,roumestan2024,fatkhiev2026monolithic}. Digital recovery must nevertheless correspond to a physical, unitary transformation to remain consistent with the security model, since a general equalizer can underestimate the excess noise \cite{fan2026practical}. Unitary compensation can reverse unitary mode mixing, although imperfect estimation and departures from a unitary channel model can leave residual coupling. Conventional single-mode security analyses treat this residual coupling as excess noise, omitting the corresponding source--receiver correlations and reducing the secret key rate. 
For DP CV-QKD, this links three requirements: parallel low-noise detection, digital recovery consistent with the security model, and key-rate estimation that accounts for incomplete mode decoupling. Their joint implementation must also accommodate time-varying coupling in deployed fibre.

Here we introduce a scalable multimode CV-QKD architecture enabled by three elements: CMOS-compatible photonic--electronic integration for parallel detection, pilot-assisted unitary digital recovery of the coupled modes, and a multimode security analysis that accounts for residual mode coupling. We implement the architecture with polarization multiplexing as a two-mode example. A silicon-photonic transmitter prepares two independently Gaussian-modulated polarization modes, while a silicon-photonic receiver co-integrated with a dual-channel 28-nm CMOS transimpedance amplifier measures the received modes in parallel, with shot-noise clearances exceeding 20~dB in both channels. From the pilot tones measured in both receiver channels, we estimate the polarization transformation and digitally compensate its unitary part, removing the need for active optical polarization control. Rather than requiring complete mode decoupling, the multimode security analysis retains the residual source--receiver correlations and incorporates them into asymptotic and finite-size key-rate estimates under collective attacks. In simulations with 2\% residual crosstalk, the multimode analysis extends the secure distance from 39~km under single-mode analysis to 84~km, near the crosstalk-free limit of 89~km. Over 10~km of deployed fibre, the mean total finite-size secret-key rate estimate is 6.69~Mbit/s, representing a 12\% increase over conventional single-mode analyses of the same digitally recovered data.

\begin{figure*}
    \centering
    \includegraphics[width=1\linewidth]{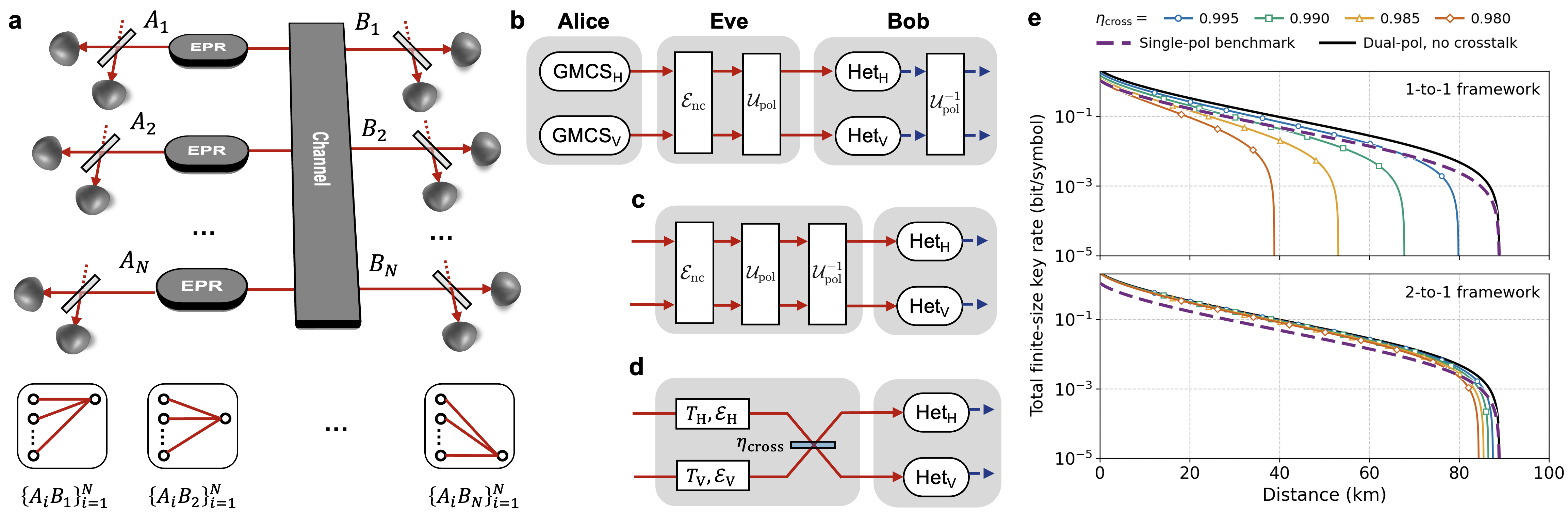}
\caption{\textbf{Multimode security framework, dual-polarization CV-QKD protocol and finite-size key rates.}
\textbf{a,} Entanglement-based scheme of an $N$-to-$N$ Gaussian-modulated coherent-state (GMCS) CV-QKD network. Grouping the links by receiver mode $B_j$ yields one $N$-to-1 subnetwork per receiver mode, in which all sources contributing to $B_j$ are retained.
\textbf{b,} Dual-polarization (DP) CV-QKD with digital polarization recovery. Alice prepares independent GMCS signals in the orthogonal H- and V-polarized modes. Eve's channel is decomposed as $\mathcal{E}=\mathcal{U}_{\mathrm{pol}}\circ\mathcal{E}_{\mathrm{nc}}$, where $\mathcal{U}_{\mathrm{pol}}$ is the unitary polarization rotation and $\mathcal{E}_{\mathrm{nc}}$ is the non-compensable component. Bob digitally applies $\mathcal{U}_{\mathrm{pol}}^{-1}$ to the heterodyne outcomes.
\textbf{c,} Equivalent representation in which $\mathcal{U}_{\mathrm{pol}}^{-1}$ is applied before heterodyne detection and absorbed into Eve's channel, valid for matched receiver branches.
\textbf{d,} Two-input, two-output model of the effective channel $\mathcal{E}_{\mathrm{nc}}$ used for the simulations. A beam splitter with transmittance $\eta_{\mathrm{cross}}$ models the residual polarization crosstalk.
\textbf{e,} Simulated finite-size secret-key rates versus distance for $\eta_{\mathrm{cross}}=0.980$--$0.995$, from the single-mode (upper) and multimode (lower) analyses. Parameters: $V_M=4$~SNU, $\varepsilon=0.02$~SNU (channel-input referred), $\beta=0.956$, $N_{\mathrm{rec}}=10^{9}$, fibre loss 0.2~dB/km and ideal detectors.}
    \label{fig:theory_foundation}
\end{figure*}

\section*{Results}
\subsection*{Multimode  security framework}


\textcolor{black}{In a multimode CV-QKD system, mode coupling introduces contributions from other independently modulated sources into each received mode. A single-mode analysis cannot distinguish these contributions from noise introduced by an eavesdropper and therefore attributes them to excess noise, making the key rate highly sensitive to residual crosstalk \cite{usenko2026continuous} and placing stringent isolation requirements on mode demultiplexing. Retaining the corresponding source--receiver correlations in a multimode description allows these signal contributions to be explicitly accounted for when bounding Eve's information, without treating the channel coupling as trusted noise. This can tighten the security bound and improve tolerance to residual mode coupling.}

We begin with an $N$-to-$N$ multi-source CV-QKD network, as illustrated in Fig.~\ref{fig:theory_foundation}a. It consists of $N$ independent EPR states whose two sets of modes, $\{A_i\}_{i=1}^{N}$ and $\{B_i\}_{i=1}^{N}$, are measured by heterodyne detection. Before detection, the $B$ modes are mixed through an $N$-input/$N$-output channel controlled by Eve, which may introduce loss, noise, and inter-mode correlations. This entanglement-based model is equivalent to a prepare-and-measure network with $N$ Gaussian-modulated coherent-state (GMCS) sources. We assume that all $N$ sources are prepared independently by trusted parties and that their modulation records are available for joint parameter estimation. We consider parallel key generation for the $N$ mutually intended mode pairs $(A_i,B_i)$. Under reverse reconciliation, the raw key for pair $j$ is defined by the measurement outcomes of $B_j$. We therefore analyse each receiver mode jointly with all source modes, forming an $N$-to-1 subnetwork ${\mathbf{A}B_j}$. 

For each receiver mode $B_j$, we consider the reduced $(N+1)$-mode state
$\rho_{\mathbf{A}B_j}$, where $\mathbf{A}=(A_1,\ldots,A_N)$, and let $E_j$
purify this state. The asymptotic secret-key
rate for the mode pair $(A_j,B_j)$ against collective attacks is
\begin{equation} \label{eq:asym_key_rate}
R_j^\mathrm{asym}=\beta_j I(a_j:b_j)-\chi(b_j:E_j)_{\rho^{G}_{\mathbf{A}B_j}},
\end{equation}
where $\beta_j$ is the reconciliation efficiency, $(a_j,b_j)$ are the heterodyne detection outcomes of $(A_j,B_j)$, $I(a_j:b_j)$ is the
classical mutual information, and $\chi(b_j:E_j)_{\rho^{G}_{\mathbf{A}B_j}}$ bounds the information available to Eve about $b_j$. Here, $\rho^{\mathrm{G}}_{\mathbf{A}B_j}$ is the Gaussian state with the same
covariance matrix $\gamma_{\mathbf{A}B_j}$ as
$\rho_{\mathbf{A}B_j}$. The optimality of Gaussian attacks gives
$\chi(b_j:E_j)_{\rho_{\mathbf{A}B_j}}
\leq
\chi(b_j:E_j)_{\rho^{\mathrm{G}}_{\mathbf{A}B_j}}$, and hence \eqref{eq:asym_key_rate} gives a lower bound on the asymptotic secret-key
rate \cite{navascues2006optimality}. The mutual information term uses only the outcomes of the intended pair, $(a_j,b_j)$, whereas the Holevo bound incorporates correlations between $B_j$ and all source modes through $\gamma_{\mathbf{A}B_j}$. Consequently, any improvement relative to the single-mode analysis of the same data arises from a tighter bound on Eve's information, while the mutual information and reconciliation efficiency remain unchanged.

In the finite-size regime, the Holevo information is evaluated using a worst-case covariance matrix $\gamma_{\mathbf{A}B_j}^{\mathrm{wc}}$ that accounts for statistical uncertainty in parameter estimation. For each quadrature, the $N$-to-1 channel is described by the Gaussian linear model $y=\mathbf{t}\mathbf{x}+z$, where $\mathbf{x}=[x_1,\ldots,x_N]^{\mathsf{T}}$ contains the independent source modulations, $\mathbf{t}=[t_1,\ldots,t_N]$ contains the coupling coefficients, and $z$ denotes residual Gaussian noise. Given the source modulation variances and receiver calibration, these $N$ transmission coefficients and the output-referred excess noise for each quadrature determine $\gamma_{\mathbf{A}B_j}$. Multivariate linear regression provides a joint confidence region for the coupling vector and an upper confidence bound on the channel-output-referred excess noise. With the excess noise fixed at its upper bound, the coupling vectors are jointly optimized within their confidence regions to maximize the Holevo information. Only candidates yielding a physically valid covariance matrix are retained, and the matrix maximizing the Holevo information defines $\gamma_{\mathbf{A}B_j}^{\mathrm{wc}}$.  Including the finite-key correction, we obtain 
\begin{equation}
    R_j^{\mathrm{finite}}
    =
    \beta_j I(a_j:b_j)
    -\chi(b_j:E_j)_{\gamma_{\mathbf{A}B_j}^{\mathrm{wc}}}
    -\Delta_{N_{\mathrm{rec}}},
\label{eq:finite-size_Rate}
\end{equation}
where $\Delta_{N_{\mathrm{rec}}}$ accounts for the finite-key penalties associated with smooth-entropy estimation and privacy amplification, and $N_{\mathrm{rec}}$ is the number of reconciled symbols for key generation. The explicit expression and security parameters are given in Methods.

Because $E_j$ purifies $\rho_{\mathbf{A}B_j}$, it includes all systems outside the subnetwork $\mathbf{A}B_j$, including the other modes
$B_{k\neq j}$. The resulting bound on $K_j$ therefore accounts for channel-induced correlations between different mode pairs. Applying the same receiver-centered analysis to every $B_j$ gives an individual key-rate bound for each intended pair. With separate reconciliation and privacy amplification, the mode pairs can generate independent secret keys in parallel. The total key rate is then $R_\mathrm{tot}=\sum_{j=1}^{N} \max\{0,R_j\}$.

\subsection*{Dual-polarization CV-QKD protocol}
The proposed security framework enables a DP CV-QKD protocol with digital polarization recovery, as illustrated in Fig.~\ref{fig:theory_foundation}b. Alice prepares Gaussian-modulated coherent states in two orthogonal polarization modes, H and V, and multiplexes them for transmission through the quantum channel. The channel is decomposed as $\mathcal{E} = \mathcal{U}_{\mathrm{pol}} \circ \mathcal{E}_{\mathrm{nc}}$, where $\mathcal{E}_{\mathrm{nc}}$ denotes the non-compensable component and $\mathcal{U}_{\mathrm{pol}}$ describes unitary polarization rotation. Here, $\circ$ denotes channel composition, with $\mathcal{E}_{\mathrm{nc}}$ acting first, followed by $\mathcal{U}_{\mathrm{pol}}$. Bob performs heterodyne detection on both received polarization modes and digitally applies the inverse transformation $\mathcal{U}_{\mathrm{pol}}^{-1}$ to the measurement outcomes. Equivalently, this inverse transformation can be applied before heterodyne detection and included in Eve's channel, as shown in Fig.~\ref{fig:theory_foundation}c. It cancels the unitary polarization rotation, leaving the effective channel $\mathcal{E}_{\mathrm{nc}}$, which may still exhibit residual polarization crosstalk. Fig.~\ref{fig:theory_foundation}d models this effective channel using polarization-dependent transmittances and excess noise, with a beam splitter of transmittance $\eta_{\mathrm{cross}}$ representing the residual crosstalk. This beam-splitter model is used only for the simulations below; in the experiment, the coupling coefficients are estimated directly without assuming a crosstalk model. The detailed asymptotic and finite-size analysis for the DP CV-QKD protocol is provided in the Methods section. 

\begin{figure*}[!t]
    \centering
    \includegraphics[width=0.98 \linewidth]{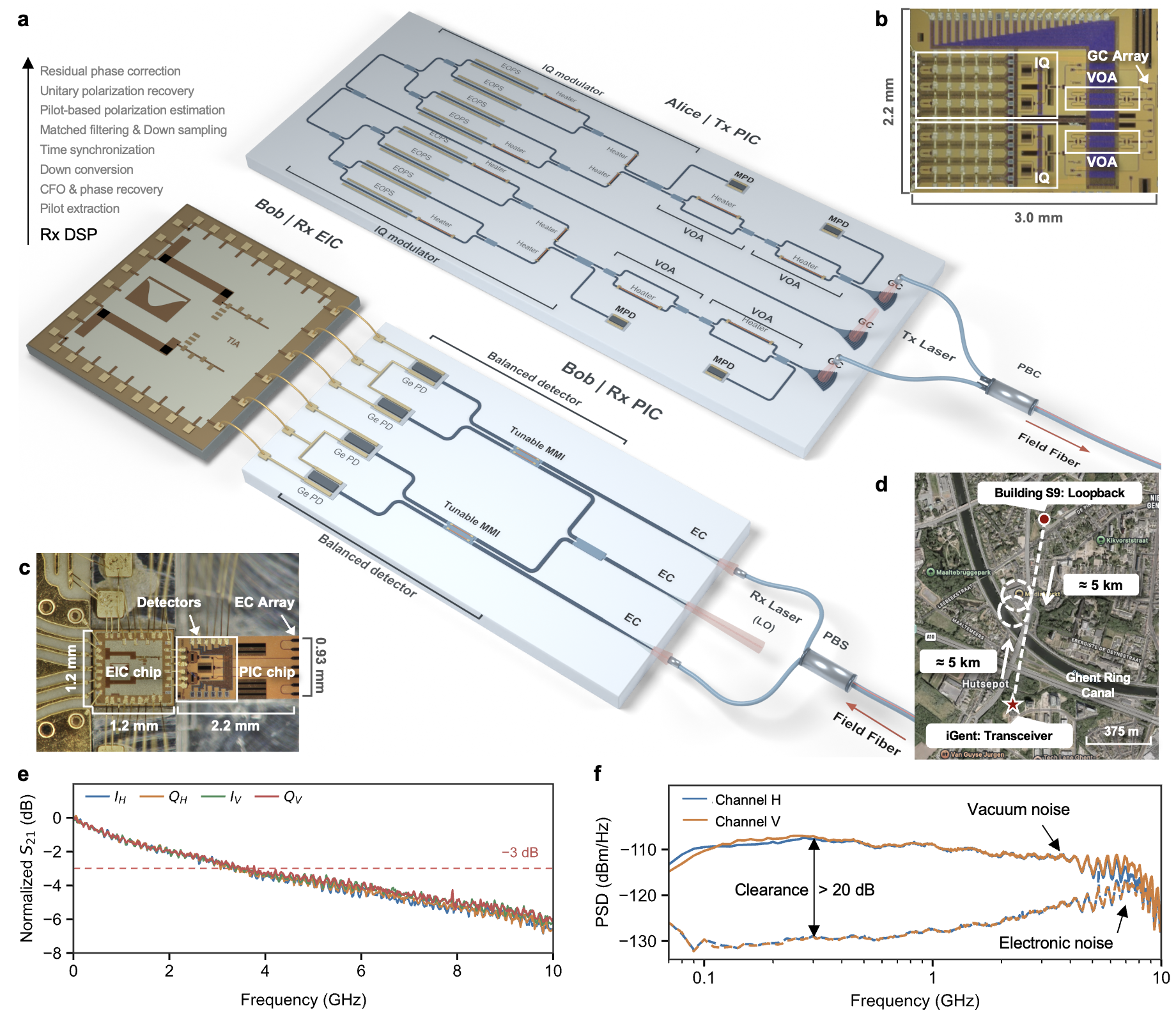}
    \caption{\textbf{Integrated dual-polarization CV-QKD system over a field fiber.} \textbf{a,} Transmitter and receiver architecture with the receiver digital signal processing (DSP) chain. \textbf{b,} Microscope photo of the transmitter PIC. \textbf{c,} Microscope photo of the receiver PIC and EIC. \textbf{d,} Route of the 10-km field-deployed fiber link. \copyright \ Apple Maps---2026. \textbf{e,} Frequency responses of the four modulation channels of the transmitter chip. \textbf{f,} Noise power spectral densities of the two receiver channels. Tx/Rx, transmitter/receiver; EIC, electronic integrated circuit; IQ, in-phase and quadrature; EOPS, electro-optic phase shifter; VOA, variable optical attenuator; MPD, monitor photodiode; GC/EC, grating/edge coupler; PBC/PBS, polarization beam combiner/splitter; MMI, multimode interferometer; Ge PD, germanium photodiode; LO, local oscillator; CFO, carrier frequency offset.
}
    \label{fig:exp_setup}
\end{figure*}

Figure~\ref{fig:theory_foundation}e compares the total finite-size secret-key rate estimates for DP CV-QKD obtained using the single-mode and multimode analyses with a single-polarization benchmark, assuming ideal detectors, $N=10^{9}$ symbols, $\beta=0.956$, $\varepsilon=0.02$~SNU and $V_M=4$~SNU. The mode-coupling-free dual-polarization benchmark is twice the single-polarization rate and shares the same maximum secure transmission distance of approximately 89 km. Under the single-mode framework, residual polarization crosstalk substantially reduces both the total key rate and the secure transmission distance. At $\eta_{\mathrm{cross}}=0.980$, the distance is limited to approximately 39 km, and the total rate remains below the single-polarization benchmark throughout the simulated range. By contrast, the multimode framework retains a total-rate advantage over single-polarization operation up to approximately 80 km at the same crosstalk level, with positive key rates extending to approximately 84 km. The single-mode analysis reaches a comparable distance only for $\eta_{\mathrm{cross}}=0.995$, corresponding to a fourfold tolerance to residual crosstalk. The impact of residual crosstalk also depends on the modulation variance: increasing the modulation in one polarization to facilitate reconciliation increases the equivalent excess noise in the orthogonal polarization under a single-mode description. Supplementary Material~S1-A and Figs.~S2 and S3 detail this trade-off and show how retaining cross-polarization correlations in the multimode analysis mitigates the associated key-rate penalty.

\subsection*{Secure digital polarization recovery}
Under the proposed framework, secure digital polarization recovery requires a unitary digital operation, as it can be moved before detection and absorbed into Eve's channel without altering the security analysis. However, polarization-dependent loss can make the $2\times2$ channel transformation non-unitary. Therefore, directly inverting the estimated channel may yield a non-unitary digital equalizer that compromises security  \cite{fan2026practical}.
To suppress residual polarization-induced crosstalk while satisfying this unitarity requirement even under non-unitary channel conditions, we develop a pilot-assisted unitary digital polarization recovery method.
We first estimate the polarization mixing matrix from the relative
complex responses of the two pilot tones across the receiver branches:
\begin{equation}
    \mathbf{H}_{\mathrm{pol}}
    =[\mathbf{u}_{\mathrm{H}},\mathbf{u}_{\mathrm{V}}],
    \quad
    \mathbf{u}_{j}
    =\frac{(1,r_{j})^{\mathrm{T}}}{\sqrt{1+|r_{j}|^{2}}},
    \quad
    r_{j}=\frac{p_{2,j}}{p_{1,j}}.
\end{equation}
Here, $j\in\{\mathrm{H},\mathrm{V}\}$, and $p_{i,j}$ denotes
the complex response of pilot $j$ in receiver branch $i$. $r_j$ is the complex response ratio between the two receiver branches for pilot $j$, encoding their relative amplitude and phase, and $\mathbf{u}_j$ is the corresponding unit-norm polarization response vector.
To retain only the unitary component of the estimated transformation, we project $\mathbf{H}_{\mathrm{pol}}$ onto its nearest unitary matrix in the Frobenius norm using singular-value decomposition:
\begin{equation}
    \mathbf{H}_{\mathrm{pol}}
    =\mathbf{U}\mathbf{\Sigma}\mathbf{V}^{\dagger},
    \qquad
    \mathbf{Q}_{\mathrm{pol}}
    =\mathbf{U}\mathbf{V}^{\dagger}.
\end{equation}
Here, $\mathbf{U}$ and $\mathbf{V}$ are unitary matrices whose columns are the left and right singular vectors of $\mathbf{H}_{\mathrm{pol}}$, respectively; $\mathbf{\Sigma}$ is a diagonal matrix containing its non-negative singular values. The resulting matrix $\mathbf{Q}_{\mathrm{pol}}$ is the unitary approximation used for polarization recovery.

The H- and V-polarized quantum signals are then recovered by applying the Hermitian transpose of this unitary matrix to the noise-normalized receiver outputs:
\begin{equation}
    \begin{bmatrix}
        \widehat{b}_{\mathrm{H}} \\
        \widehat{b}_{\mathrm{V}}
    \end{bmatrix}
    =
    \mathbf{W}_{\mathrm{pol}}
    \begin{bmatrix}
        \widetilde{b}_{1} \\
        \widetilde{b}_{2}
    \end{bmatrix},
    \qquad
    \mathbf{W}_{\mathrm{pol}}
    =\mathbf{Q}_{\mathrm{pol}}^{\dagger}
    =\mathbf{V}\mathbf{U}^{\dagger}.
\end{equation}
The same unitary recovery matrix is applied to the corresponding noise-calibration data, ensuring that the calibration describes the same recovered modes as the signal measurement. 
The unitary digital recovery is equivalent to a physical polarization transformation and suppresses polarization-induced crosstalk. After the polarization recovery, correlations arising from residual crosstalk are incorporated into the security analysis, mitigating the associated key-rate penalty.

Although demonstrated for two polarization modes, the pilot-assisted unitary recovery extends in form to an $N$-mode linear mixing channel. Distinguishable pilots and mode-resolved detection can provide an $N\times N$ mixing estimate, whose nearest unitary matrix is obtained through the same singular-value-decomposition projection and can be used to suppress the inter-mode crosstalk.

\subsection*{CMOS-compatible photonic--electronic transmitter and receiver}

Figure~\ref{fig:exp_setup}a shows the transmitter and receiver architecture. Both photonic integrated circuits (PICs) are fabricated on IMEC's silicon-photonics platform, and the receiver PIC is co-integrated with a dual-channel transimpedance amplifier (TIA) fabricated in 28-nm CMOS. The PICs prepare and detect two parallel optical channels, while external multiplexing and demultiplexing components map these channels onto the transmission modes. Here, a polarization beam combiner at the transmitter and a polarization beam splitter at the receiver implement polarization multiplexing. This separation between state preparation and detection and the external mode mapping provides a basis for adapting the architecture to other optical degrees of freedom, using suitable spatial-mode interfaces or wavelength-selective components with appropriate carrier and local-oscillator arrangements.

The transmitter PIC (Fig.~\ref{fig:exp_setup}a and b) divides the optical carrier into two branches using a $1\times2$ multimode interferometer (MMI). The mean insertion losses of the $1\times2$ and $2\times2$ MMIs are 0.11 and 0.21~dB, respectively. In each branch, two travelling-wave Mach--Zehnder modulators driven in push--pull form an IQ modulator \cite{cammarata2026twmzm}, followed by two cascaded thermo-optic variable optical attenuators (VOAs) that reduce the signal power to the quantum level. Monitor photodiodes at the IQ-modulator output and after the second VOA support automatic bias control and power monitoring. To reduce thermal crosstalk between the two channels on the shared die, the bias heaters of adjacent modulators and attenuators are placed on opposite interferometer arms, increasing their physical separation. Light is coupled into and out of the transmitter PIC through grating couplers, to which a fibre array is fixed with adhesive. The four modulators in the packaged transmitter exhibit matched electro-optic responses with a 6-dB bandwidth of approximately 9~GHz, including the effects of the printed-circuit-board traces and radio-frequency cables (Fig.~\ref{fig:exp_setup}e). This bandwidth exceeds that required at the present symbol rate. 

The receiver PIC (Fig.~\ref{fig:exp_setup}a and c) accepts the two signal outputs of the off-chip polarization beam splitter and the local oscillator (LO) through edge couplers, which are interfaced with an actively aligned fibre array that is not permanently attached. A $1\times2$ MMI divides the LO between the two detection channels. In each channel, a tunable $2\times2$ MMI coupler combines the signal and LO and enables balancing of the optical power incident on a photodiode pair with a quantum efficiency of 86.5\%. Each balanced photodiode pair is wire-bonded to one channel of the CMOS TIA (Fig.~\ref{fig:exp_setup}c), giving a receiver footprint of $1.2\times3.4$~mm$^2$.

Achieving low-noise amplification in deep-submicron CMOS requires addressing the noise of the input transistors and feedback network. The TIA uses a capacitive-feedback topology that replaces the conventional feedback resistor with a capacitor, eliminating the thermal-noise contribution of that feedback resistor. A resistor and a capacitor ratio set the transimpedance, allowing gain and bandwidth to be controlled more independently and reducing sensitivity to process variations. An on-chip DC-cancellation loop prevents residual photodiode imbalance from saturating the integrating front end \cite{TIA_sarah}. The TIA consumes 50~mW across both channels.

The two receiver channels achieve shot-noise clearances of 21.1 and 22.0~dB at 0.3~GHz (Fig.~\ref{fig:exp_setup}f), with both exceeding 20~dB from approximately 0.1 to 1~GHz. These measurements use LO photocurrents of 680 and 575~$\mu$A, respectively. For comparison, a standalone detector using the same TIA design achieved a clearance of 27~dB at a photocurrent of 2.2~mA \cite{TIA_sarah}. The lower LO photocurrents used here reduce the shot-noise power and therefore contribute to the lower measured clearances.

\begin{figure*}[!t]
    \centering
    \includegraphics[width=0.9\linewidth]{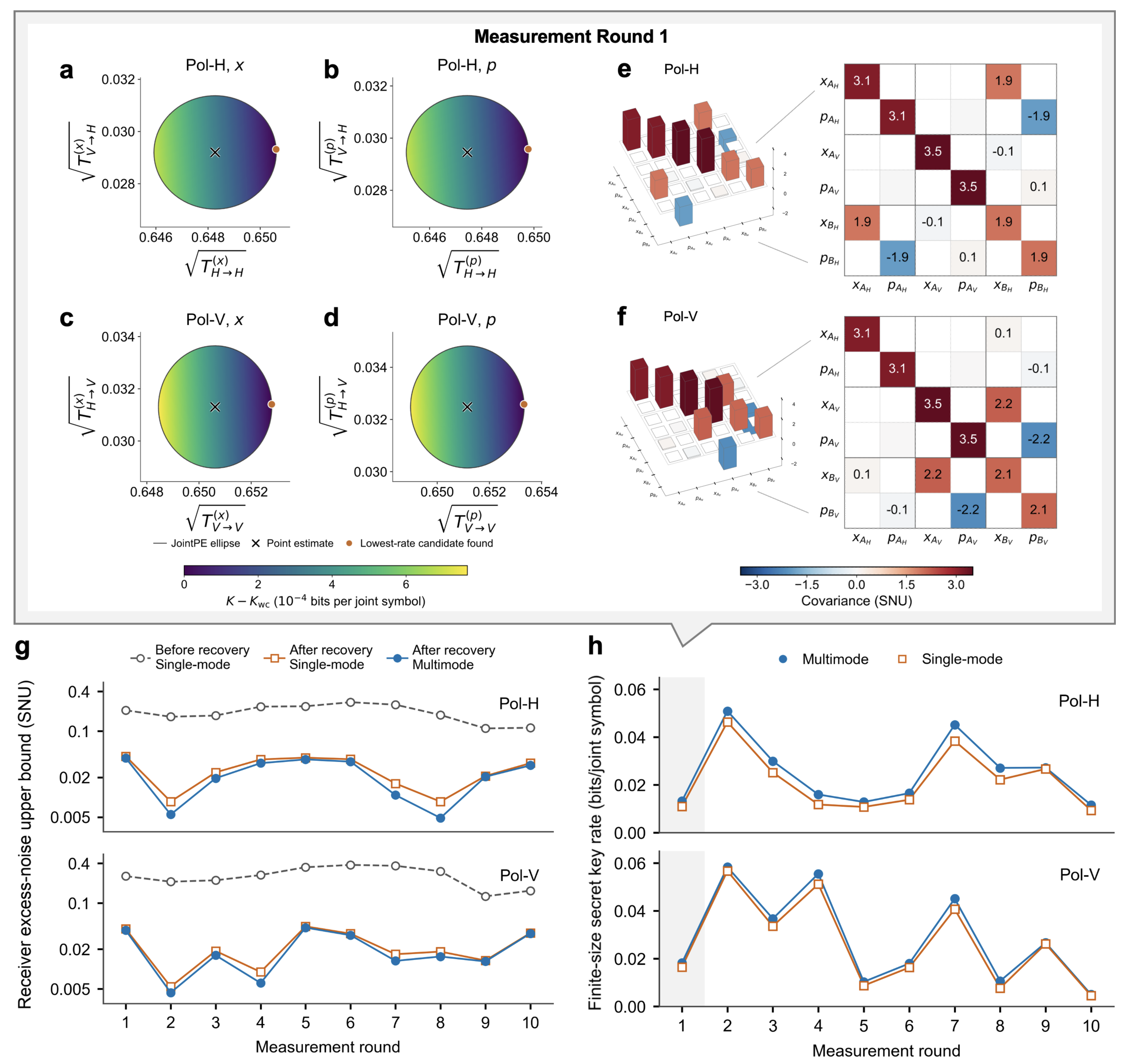}
    \caption{\textbf{Joint parameter estimation and finite-size key rates in the field implementation.}
\textbf{a--d,} Key-rate maps within the joint transmittance confidence ellipses for the $x$ and $p$ quadratures of Pol-H (\textbf{a,b}) and Pol-V (\textbf{c,d}) in measurement round~1. Crosses indicate point estimates; orange circles mark the lowest-rate candidate found by jointly optimizing the four coupling coefficients of each polarization to maximize Eve's Holevo information. Each map is a two-dimensional slice with the other quadrature's transmittances fixed at the selected values and excess noises at their upper confidence bounds. Colors show $K-K_{\mathrm{wc}}$, relative to the corresponding worst-case finite-size key rate.
\textbf{e,f,} Worst-case three-mode covariance matrices for Pol-H (\textbf{e}) and Pol-V (\textbf{f}) from the finite-size analysis.
\textbf{g,} Receiver-referred excess-noise upper bounds at the channel output over ten measurement rounds for the H and V polarizations, averaged over the separately evaluated $x$- and $p$-quadrature bounds. Curves compare the conventional single-mode analysis before and after digital polarization recovery with the multimode analysis after recovery. The vertical axis is logarithmic.
\textbf{h,} Corresponding finite-size secret key rates from the multimode and single-mode analyses of the same data after digital polarization recovery.
Shading marks round~1, detailed in \textbf{a--f}. Both analyses use the same experimental and security parameters; calculation details are given in Supplementary Section~S1.}
    \label{fig:exp_results}
\end{figure*}

\subsection*{Field implementation}

We operate the system over a deployed campus fibre loop at Technologiepark Zwijnaarde in Ghent, Belgium (Fig.~\ref{fig:exp_setup}d). The transmitter and receiver are both located in the iGent building; the fibre runs to a loopback point at Building~S9 of Ghent University, crossing the Ghent ring canal twice, and returns to iGent. The round-trip length is 10~km, with a channel loss of approximately 4~dB. Because the link is a loop, the transmitter and receiver are co-located and share a common clock reference, whereas the LO is generated by an independent laser at the receiver. The system structure is detailed in Supplemental Material S2.
 
Each IQ modulator on the transmitter chip generates a single-sideband, Gaussian-modulated quantum signal at a symbol rate of 125~MBaud, centred at 250~MHz from the carrier, with a modulation variance of 2.1 and 2.5~SNU for polarization H and V, respectively. A pilot tone is frequency-multiplexed with each quantum signal, at 100~MHz for Pol-H and 120~MHz for Pol-V. The drive signals are generated by two arbitrary waveform generators (AWGs) that share a clock and trigger with the analog-to-digital converter (ADC). 
The polarization beam combiner places the two transmitter outputs onto orthogonal polarization modes of the fibre, and the polarization beam splitter at the channel output separates the received light into two orthogonal modes. No polarization controller is used, so each received mode is a time-varying mixture of both transmitted modes, determined by the polarization rotation accumulated in the deployed fibre. On the receiver chip, each mode is combined with the local oscillator and measured by radio-frequency heterodyne detection. To increase the electrical output swing, the TIA outputs are amplified using external RF amplifiers (SHF S807). The two TIA outputs are then digitized by a 16-bit ADC at 1~GSa/s.

The two digitized channels are processed jointly (Fig.\ref{fig:exp_setup}a). Pilot-assisted frequency and phase recovery removes the frequency offset and relative phase drift between the transmitter laser and the local oscillator. The quantum signals are then digitally down-converted to baseband, temporally synchronized, filtered with a root-raised-cosine matched filter and sampled at the symbol rate. Because both pilot tones appear in both receiver channels, their relative amplitudes and phases determine the polarization transformation, which is compensated by the pilot-assisted unitary recovery described above, without active optical polarization control. The allocation of pilot signals in the frequency domain is detailed in Supplemental Material S2.

Each measurement round contains $N=2.5\times10^{7}$ symbols per polarization, and ten rounds were recorded randomly over several hours. We use the one-time calibration scheme, in which electronic noise is mapped to an equivalent optical loss \cite{zhang2020one}. The maximum overall detection efficiency of the two receiver channels, including the polarization beam splitter, edge-coupling and on-chip losses and the photodiode efficiency, is calibrated to $\eta_d=0.334$ and treated as trusted. Using the maximum detection efficiency to define the trusted receiver loss preserves the symmetry between the two detector models while maintaining a conservative security analysis, as any additional detection loss is attributed to Eve.

\subsection*{Field test results}

We evaluate the H- and V-polarization key rates over ten measurement rounds after pilot-assisted unitary polarization recovery. Each round is analysed using both the multimode and conventional single-mode frameworks applied to the same recovered data. The multimode analysis evaluates the Pol-H key rate using the three-mode Gaussian state of $A_HA_VB_H$, and the Pol-V key rate using that of $A_HA_VB_V$. The single-mode analysis uses only $A_HB_H$ or $A_VB_V$, treating the contribution from the other source as excess noise. Both analyses use the same mutual information between each receiver and its intended source, receiver calibration, reconciliation efficiency $\beta=0.96$, and finite-size parameters. They differ only in the bound on Eve's Holevo information, isolating the effect of retaining residual source--receiver correlations. Rates are expressed in bits per joint symbol, where one joint symbol comprises one transmitted symbol on each polarization.

For finite-size parameter estimation in the multimode analysis, we characterize each receiver mode by its direct and cross transmittances and receiver-referred excess noise in both quadratures. For Pol-H, these parameters are $T_{H\rightarrow H}^{(q)}$, $T_{V\rightarrow H}^{(q)}$ and $\xi_H^{(q)}$, with $q\in\{x,p\}$; the parameters for Pol-V follow by exchanging H and V. For each round and polarization, we fix the excess noise of both quadratures at its upper confidence bound and jointly optimize the four coupling coefficients within their confidence regions to maximize Eve's Holevo information. Only physically valid covariance matrices are retained, and the lowest finite-size key rate found is denoted $K_{\mathrm{wc}}$. Figure~\ref{fig:exp_results}a--d shows slices through the resulting key-rate landscape for round~1. In each plotted slice, the minimizing candidate lies near the largest direct transmittance permitted by the confidence ellipse. Within these regions, increasing the direct transmittance increases the evaluated Holevo information and therefore lowers the key rate, since the mutual information is held fixed. The rate varies by less than $8\times10^{-4}$~bits per joint symbol across the plotted regions, corresponding to approximately 6\% of the round-1 Pol-H rate.

Figure~\ref{fig:exp_results}e,f shows the corresponding worst-case covariance matrices. Each recovered receiver mode remains correlated with both sources. For Pol-H, the correlation of $B_H$ with its intended source is 1.9~SNU, while its correlation with the orthogonal source has a magnitude of 0.1~SNU. The worst-case transmittances give residual cross-to-direct transmittance ratios, averaged over the $x$ and $p$ quadratures, of $T_{V\rightarrow H}/T_{H\rightarrow H}\approx\ 2.05\times10^{-3}$ ($-26.9$~dB) for Pol-H and $T_{H\rightarrow V}/T_{V\rightarrow V}\approx 2.40\times10^{-3}$ ($-26.2$~dB) for Pol-V. The single-mode analysis treats these contributions as excess noise, whereas the multimode analysis explicitly retains their correlations with the other source. The direct and cross transmittances are jointly estimated from the modulation and measurement records, without assuming that the channel crosstalk is trusted.

Figure~\ref{fig:exp_results}g shows the excess-noise upper bounds from the conventional single-mode analysis before and after digital polarization recovery and the multimode analysis after recovery. In the single-mode analysis, the excess noise is the residual variance above the calibrated vacuum baseline after accounting only for the modulation of the intended source; it therefore includes the signal contribution from the orthogonal polarization. In the multimode analysis, both source modulations are included in the estimation (see \eqref{eq} in Methods), so the excess noise is evaluated relative to the same calibrated vacuum baseline after accounting for both signal contributions. Each point is the mean value of the separately evaluated $x$- and $p$-quadrature upper bounds, plotted on a logarithmic vertical axis.

Before recovery, the single-mode excess-noise upper bounds range from approximately 0.110 to 0.378~SNU. Digital polarization recovery reduces these bounds to below 0.0442~SNU. By distinguishing residual cross-polarization signal contributions from excess noise, the multimode analysis further lowers the bounds by an average of $3.20\times10^{-3}$~SNU for Pol-H and $1.93\times10^{-3}$~SNU for Pol-V over the ten rounds, while retaining the associated correlations in the evaluation of Eve's Holevo information. The simulations in Supplementary Section~S1A further illustrate that residual crosstalk can substantially increase the equivalent excess noise under single-mode analysis, while its key-rate penalty is substantially reduced when the associated source--receiver correlations are retained. This enhanced tolerance to residual coupling may relax the precision required for digital polarization recovery, allowing lower-complexity processing with reduced computational overhead.

The finite-size key-rate estimates in Fig.~\ref{fig:exp_results}h broadly follow these variations: low-noise rounds, such as round~2, yield higher rates, whereas high-noise rounds, such as round~5, yield lower rates. Both polarizations have positive key-rate estimates in every round, and the multimode estimate is higher than the single-mode estimate throughout. The absolute increase ranges from approximately $4.56\times10^{-4}$ to $6.82\times10^{-3}$~bits per joint symbol. The relative increase ranges from below 2\% for Pol-V in round~9 to approximately 38.9\% for Pol-V in round~8. In round~5, the relative improvements are approximately 19.3\% and 18.3\% for Pol-H and Pol-V, respectively.

Averaged over the ten rounds, the H-polarization key-rate estimate increases from 0.02153 to 0.02504~bits per joint symbol, and the V-polarization estimate increases from 0.02620 to 0.02845~bits per joint symbol. These correspond to relative improvements of 16.3\% and 8.6\%, respectively. The total increases from 0.04774 to 0.05349~bits per joint symbol, giving an overall improvement of 12\%, calculated from the ratio of the two mean total rates. At 125~MBaud per polarization, the multimode estimates correspond to {3.13}~Mbit/s for H and {3.56}~Mbit/s for V, giving a mean total of {6.69}~Mbit/s, compared with {5.97}~Mbit/s under the single-mode analysis. These are finite-size secret-key rate estimates against collective attacks; error correction and privacy amplification are not executed in this work.

The comparison demonstrates the complementary roles of digital recovery and multimode security analysis. In the round-1 worst-case estimates, unitary recovery leaves {quadrature-averaged} cross-to-direct transmittance ratios of approximately {$-26.9$~dB for Pol-H and $-26.2$~dB for Pol-V} without active optical polarization control. Retaining the remaining source--receiver correlations then tightens the bound on Eve's information and increases the estimated key rate without changing the hardware or digital recovery.

\arrayrulecolor{black}
\begin{table*}[t]
    \centering
    \caption{Comparison of chip-based CV-QKD demonstrations.
    $L_{\max}$ denotes the maximum reported distance, and SKR represents the total secret key rate per transmitted symbol, where a single symbol consists of two orthogonal polarization components in DP configuration. $^*$, simulated results; $^\dagger$, emulated results via VOA in place of a fiber channel; $^\ddagger$, finite-size channel estimation without full finite-size security proof.}
    \label{tab:chip_cvqkd_comparison}
    \small
    \setlength{\tabcolsep}{3pt}
    \renewcommand{\arraystretch}{1.2}
    \arrayrulecolor{black}

    \begin{tabular*}{\linewidth}{@{}lllllllll@{}}
        \toprule
        Reference & Setting & Protocol & Tx & Rx
        & Amplifier technology & PC & Regime & SKR at $L_{\max}$ (bit/symbol) \\
        \midrule

        Zhang et al.\ (2019)~\cite{zhang2019integrated}
        & Lab & Single-pol. Gaussian & Si PIC & Si PIC
        & — & Bulk & Asymptotic
        & $1.75\times10^{-4}$ @ 100~km $^*$ \\

        \addlinespace
        Bian et al.\ (2024)~\cite{bian2024continuous}
        & Lab & Single-pol. Gaussian & Bulk & Si PIC
        & — & Bulk & Finite-size
        & $2.40\times10^{-4}$ @ 28.6~km \\

        \addlinespace
        Hajomer et al.\ (2024)~\cite{hajomer2024continuous}
        & Lab & Single-pol. 16-QAM & Bulk & Si PIC
        & GaAs pHEMT & Bulk & Finite-size
        & 0.0351 @ 10~km \\

        \addlinespace
        Pi\'etri et al.\ (2024)~\cite{pietri2024experimental}
        & Lab & Single-pol. Gaussian & Bulk & Si PIC
        & Commercial & N/A & Asymptotic
        & 0.00220 @ 23~km $^\dagger$ \\

        \addlinespace
        Aldama et al.\ (2025)~\cite{aldama2025integrated}
        & Lab & Single-pol. Gaussian & InP PIC & Bulk
        & Commercial & Bulk & Asymptotic
        & 0.00488 @ 11~km \\

        \addlinespace
        Hajomer et al.\ (2026)~\cite{hajomer2025chip}
        & Lab & Single-pol. 16-QAM & Si PIC & Si PIC
        & GaAs pHEMT & Bulk & Finite-size
        & 0.0154 @ 20~km \\

        \addlinespace
        Ng et al.\ (2026)~\cite{ng2025gigabit}
        & Lab & Dual-pol. QPSK & Si PIC & Si PIC
        & Commercial & Bulk & Finite-size
        & 0.0303 @ 10~km \\

        \addlinespace
        Servello et al.\ (2026)~\cite{servello2026chip}
        & Lab & Single-pol. Gaussian & Si PIC & Bulk
        & — & Digital & Asymptotic
        & $1.02\times10^{-5}$ @ 90~km $^\dagger$\\

        \addlinespace
        Konig et al.\ (2026)~\cite{konig2026system}
        & Lab & Single-pol. 64-QAM & InP PIC & Bulk
        & Commercial & Digital & Asymptotic
        & 0.055 @ 14.3~km \\

        \addlinespace
        Xu et al.\ (2026)~\cite{xu2026fully}
        & Lab & Single-pol. Gaussian & Si PIC & Si PIC
        & — & Bulk & Finite-size
        & $1.29\times10^{-4}$ @ 126.06~km \\

        \addlinespace
        Fatkhiev et al.\ (2026)~\cite{fatkhiev2026monolithically}
        & Lab & Dual-pol. PS-64-QAM & Si PIC & Si PIC
        & Co-packaged TIA & Digital & Finite-size\textsuperscript{$\ddagger$}
        & 0.00760 @ 25.2~km \\

        \addlinespace
        \textbf{This work}
        & \textbf{Field}
        & \textbf{Dual-pol. Gaussian}
        & \textbf{Si PIC}
        & \textbf{Si PIC}
        & \textbf{28 nm CMOS}
        & \textbf{Digital}
        & \textbf{Finite-size}
        & \textbf{0.0535 @ 10~km} \\

        \bottomrule
    \end{tabular*}
\end{table*}

\section*{Discussion}

We have shown that the two routes to scaling CV-QKD, multimode transmission and photonic--electronic integration, can be combined in a single system operated over deployed fibre. Silicon-photonic transmitter and receiver chips with a dual-channel 28-nm CMOS TIA prepare and detect two polarization modes in parallel, pilot-assisted unitary recovery replaces active optical polarization control, and the multimode security analysis retains the residual coupling that recovery leaves behind. The two steps act in sequence: unitary recovery alone reduces the single-mode excess-noise bounds from 0.110--0.378~SNU to below 0.044~SNU, and the multimode analysis then removes most of the key-rate penalty associated with the remaining coupling. Over 10~km of deployed fibre, the system yields a mean total finite-size key-rate estimate of 0.05349~bits per joint symbol, or 6.69~Mbit/s, which is 12.1\% higher than the conventional single-mode estimate from the same recovered data. This improvement requires no additional hardware or signal processing; it follows only from how the measured data are analysed.
 
Table~\ref{tab:chip_cvqkd_comparison} places these results among chip-based CV-QKD demonstrations. The present system is the only one operated over deployed fibre, the only one with multichannel CMOS electronic readout, and, together with Ref.~\citenum{fatkhiev2026monolithically}, one of two dual-polarization chip-based implementations with fully digital polarization recovery. In Ref.~\citenum{fatkhiev2026monolithically}, polarization is recovered with a constant-modulus equalizer that is not constrained to be unitary, and each polarization is analysed separately, without accounting for residual coupling between the modes. As shown in Ref.~\citenum{fan2026practical}, non-unitary dynamic equalization can lead to an underestimation of excess noise unless its effect on shot-noise statistics is properly accounted for in the security analysis. In the present work, the digital recovery is restricted to unitary operations compatible with the security analysis, and the residual source--receiver correlations are retained in a multimode finite-size key-rate estimate. Because both polarization modes contribute key simultaneously, the total finite-size rate of 0.0535~bits per transmitted symbol is the highest among the finite-size demonstrations in Table~\ref{tab:chip_cvqkd_comparison}. Per polarization mode, the rate of approximately 0.027~bits per symbol is comparable to laboratory single-mode systems at similar distances \cite{hajomer2024continuous,hajomer2025chip}, despite operation over a deployed link with time-varying polarization and excess noise of up to 0.044~SNU.
 
The multimode security analysis has implications beyond the present experiment. In any multiplexed system, the separation of modes is limited by the finite extinction of the multiplexing components and by imperfect compensation of channel-induced coupling. A single-mode analysis converts every residual coupling into excess noise, so the tolerable crosstalk, and hence the required isolation of the mode demultiplexing, sets the achievable distance. Retaining the source--receiver correlations relaxes this requirement: in the simulations of Fig.~\ref{fig:theory_foundation}e, residual crosstalk of $\eta_{\mathrm{cross}}=0.98$ limits the single-mode analysis to approximately 39~km, whereas the multimode analysis extends the reach to approximately 84~km, close to the crosstalk-free value of approximately 89~km. The field data show the same trend: the relative improvement is largest in the rounds with the lowest single-mode rates, reaching 38.9\% for Pol-V in round~8, so the analysis is most beneficial under the least favourable channel conditions. It also decouples two design choices that a single-mode analysis ties together. Because crosstalk converts the modulation of one mode into excess noise on the other, a single-mode analysis forces the modulation variance to be chosen jointly with the isolation of the mode demultiplexing; at 1\% residual crosstalk and $V_{M}=4$~SNU, the equivalent excess noise increases approximately threefold, whereas the multimode analysis recovers most of the crosstalk-free key rate (Supplementary Section~S1A). This shifts part of the burden of mode separation from hardware and active control to the security analysis, which is particularly relevant for integrated platforms, where on-chip polarization splitters, mode multiplexers and filters have limited extinction and active control adds loss and complexity. The framework may also be extended to temporally correlated signals by treating distinct temporal modes jointly and retaining their source--receiver correlations in the security analysis.
 
Several aspects of the present demonstration leave room for further improvement. The overall detection efficiency of $\eta_d=0.334$ is primarily limited by receiver coupling, while the photodiodes have a quantum efficiency of 86.5\%; improved edge coupling and packaging would therefore increase the key rate. The symbol rate of 125~MBaud per polarization is limited by the AWG/ADC bandwidth, whereas the transmitter modulators have a 6-dB bandwidth of approximately 9~GHz, and the same TIA design has demonstrated shot-noise-limited detection up to 13.7~GHz in a standalone detector \cite{TIA_sarah}. These component bandwidths provide scope for higher-rate operation, as demonstrated for single-channel integrated CV-QKD at 10--16~GBaud \cite{hajomer2024continuous,hajomer2025chip}. Further hardware development includes integrating the lasers and polarization multiplexing components, advancing beyond wire-bonded photonic--electronic integration, and permanently attaching the receiver fibre array. On the protocol side, the reported key-rate estimates account for collective attacks and finite-size effects at an assumed reconciliation efficiency. Future work will implement error correction and privacy amplification for end-to-end key generation and extend the multimode framework to composable security.
 
The architecture extends naturally beyond two polarization modes. The photonic chips prepare and detect parallel channels independently of the optical degree of freedom, which is set by the external multiplexing components, so the same architecture can address frequency or spatial modes. For $N$ independently modulated signals, the multimode analysis uses an expanded covariance matrix that includes every source correlated with a given receiver mode, and the digital recovery generalizes from a $2\times2$ to an $N\times N$ unitary estimated from the multimode measurement outcomes. Because each receiver mode is then characterized by $N$ transmittances per quadrature, the dimension of the confidence region grows with $N$; quantifying how the finite-size penalty and the computational cost of the worst-case search scale with the number of modes is therefore a key next step. At the hardware level, the present TIA requires 25~mW per channel within a receiver footprint of $1.2\times3.4$~mm$^2$, and CMOS offers a direct route to receiver arrays with many low-noise channels and to co-integration of amplification with analog-to-digital conversion and digital signal processing on the same technology platform. Such multichannel transceivers could support multi-carrier CV-QKD \cite{wang2025high}, upstream access networks \cite{xu2025ofdm} and multichannel quantum random number generation \cite{haylock2019multiplexed}, each according to its mode-access and source-trust requirements. Together, these directions point to multimode CV-QKD systems in which parallelism, integration and security analysis are co-designed from the outset.

\section*{Methods}

\subsection*{Asymptotic security analysis of the dual-polarization protocol}

We specialize the multimode framework to the two independently
modulated polarization modes. For the H-polarization key, we retain
Alice's modes $A_H$ and $A_V$ together with Bob's recovered mode $B_H$.
All systems outside this three-mode description are conservatively
included in a purifying system $E'_H$, which includes the unused
receiver mode $B_V$. The two source modes are trusted, whereas the
channel and its residual polarization coupling are untrusted. The
analysis therefore retains the correlations associated with
cross-polarization coupling without treating this coupling as trusted
noise.

We use quadrature units in which the vacuum variance is unity.
In the ordering
$(x_{A_H},p_{A_H},x_{A_V},p_{A_V},x_{B_H},p_{B_H})$,
the three-mode covariance matrix is written as
\begin{equation}
    \gamma_{A_H A_V B_H}
    =
    \begin{pmatrix}
        \gamma_{A_H} & \mathbf{0} & C_{A_H,B_H} \\
        \mathbf{0} & \gamma_{A_V} & C_{A_V,B_H} \\
        C_{A_H,B_H}^{\mathsf T}
        & C_{A_V,B_H}^{\mathsf T}
        & \gamma_{B_H}
    \end{pmatrix}
    =
    \begin{pmatrix}
        \gamma_A & C_H \\
        C_H^{\mathsf T} & \gamma_{B_H}
    \end{pmatrix},
    \label{eq:methods_dp_covariance}
\end{equation}
where $A=(A_H,A_V)$ and
$\gamma_A=\gamma_{A_H}\oplus\gamma_{A_V}$.
The vanishing source--source block follows from the independent
preparation of the two source states. For Gaussian modulation,
$\gamma_{A_j}=(V_{M,j}+1)\mathbf{I}_2$, where $V_{M,j}$ is the
modulation variance of polarization $j\in\{H,V\}$. The blocks
$C_{A_H,B_H}$ and $C_{A_V,B_H}$ describe the direct and cross-polarization correlations, respectively.

For collective attacks, Eve's Holevo information is bounded using
the Gaussian state with covariance matrix $\gamma_{A_H A_V B_H}$. For a
purification of this Gaussian state and ideal heterodyne detection
of $B_H$, the Holevo information is
\begin{equation}
    \chi_H(\gamma_{A_H A_V B_H})
    \equiv \chi(b_H:E'_H)
    =
    S(A_HA_VB_H)-S(A_HA_V|b_H).
    \label{eq:methods_dp_holevo_entropy}
\end{equation}
Here, $S(A_HA_VB_H)$ and $S(A_HA_V|b_H)$ can be calculated with covariance matrices $\gamma_{A_H A_V B_H}$ and $\gamma_{A_H A_V|b_H}$.
The covariance matrix of Alice's two modes conditioned on Bob's
heterodyne outcome is
\begin{equation}
    \gamma_{A_H A_V|b_H}
    =
    \gamma_A
    -
    C_H
    \left(\gamma_{B_H}+\mathbf{I}_2\right)^{-1}
    C_H^{\mathsf T}.
    \label{eq:methods_dp_conditional_covariance}
\end{equation}

The asymptotic reverse-reconciliation key rate is
\begin{equation}
    R_H^{\mathrm{asym}}
    =
    \beta_H I(a_H:b_H)-\chi_H(\gamma_{A_H A_V B_H}),
    \label{eq:methods_dp_asymptotic_rate}
\end{equation}
where $\beta_H$ is the reconciliation efficiency and
$I(a_H:b_H)$ is the mutual information between the
H-polarization modulation and measurement outcomes. 
The V-polarization rate is obtained by retaining
$A_HA_VB_V$ and exchanging the polarization labels. The beam-splitter model used to simulate residual crosstalk represents a particular channel realization, whereas the three-mode security analysis is formulated in terms of the source--receiver covariance matrix, which can be estimated directly from the modulation and measurement data without assuming a specific channel structure.

\subsection*{Finite-size security analysis and joint parameter estimation}

We account for finite-sample uncertainty by constructing confidence
regions for the direct and cross-polarization coupling coefficients
and the residual-noise variance.
For each quadrature $q\in{x,p}$, Bob's H-polarization measurement
is described by the two-input linear model
\begin{equation}
b_H^q
=
t_{HH}^q a_H^q
+
t_{VH}^q a_V^q
+
z_H^q,
\label{eq}
\end{equation}
where $a_H^q$ and $a_V^q$ are Alice's Gaussian modulation variables
for the H and V polarizations, respectively, and $b_H^q$ is Bob's
corresponding H-polarization measurement outcome.
The superscript $q$ labels the $x$ or $p$ quadrature.
The coefficients $t_{HH}^q$ and $t_{VH}^q$ describe the direct
$H\rightarrow H$ and cross-polarization $V\rightarrow H$ couplings,
respectively.
The residual-noise samples $z_H^q$ are assumed to be independent and identically distributed zero-mean Gaussian variables with variance $\sigma_{z,q}^{2}$, and independent of Alice's modulation data.

For $N_{\mathrm{rec}}$ parameter-estimation samples, let
$X_q\in\mathbb{R}^{N_{\mathrm{rec}}\times 2}$ be the design matrix
whose columns contain Alice's modulation samples $a_H^q$ and
$a_V^q$, respectively, and let
$\mathbf{b}_{H,q}\in\mathbb{R}^{N_{\mathrm{rec}}}$ contain Bob's
corresponding H-polarization measurement outcomes.
Writing $\mathbf{t}_q=(t_{HH}^q,t_{VH}^q)^{\mathsf T}$ and
assuming that $X_q$ has full column rank, the least-squares
estimate of the coupling vector and the unbiased estimate of
the residual-noise variance are
\begin{equation}
    \begin{aligned}
        \widehat{\mathbf{t}}_q
        &=
        \left(X_q^{\mathsf T}X_q\right)^{-1}
        X_q^{\mathsf T}\mathbf{b}_{H,q},
        \\
        \widehat{\sigma}_{z,q}^{\,2}
        &=
        \frac{
            \left\|
                \mathbf{b}_{H,q}-X_q\widehat{\mathbf{t}}_q
            \right\|_2^2
        }{N_{\mathrm{rec}}-2}.
    \end{aligned}
    \label{eq:methods_dp_estimators}
\end{equation}
The denominator $N_{\mathrm{rec}}-2$ accounts for the two fitted
coupling coefficients in the model.
We construct a joint confidence region for the coupling vector,
with its geometry determined by $X_q^{\mathsf T}X_q$.

Let $\nu=N_{\mathrm{rec}}-2$ denote the residual degrees of
freedom. Under the Gaussian regression model specified above,
an exact joint confidence region for the two coupling
coefficients, with confidence level $1-\epsilon_{t,q}$, is
\begin{equation}
    \begin{aligned}
        \mathcal{E}_q
        =
        \Bigl\{
            \mathbf{t}_q\in\mathbb{R}^{2}:\;&
            (\mathbf{t}_q-\widehat{\mathbf{t}}_q)^{\mathsf T}
            X_q^{\mathsf T}X_q
            (\mathbf{t}_q-\widehat{\mathbf{t}}_q)
            \\
            &\le
            2\widehat{\sigma}_{z,q}^{\,2}
            F^{-1}_{2,\nu}(1-\epsilon_{t,q})
        \Bigr\},
    \end{aligned}
    \label{eq:methods_dp_confidence_region}
\end{equation}
where $F^{-1}_{2,\nu}$ is the quantile function of the
$F$ distribution with degrees of freedom $2$ and $\nu$.
A one-sided upper confidence bound on the residual-noise
variance, with confidence level $1-\epsilon_{\sigma,q}$, is
\begin{equation}
    \sigma_{z,q,\mathrm{up}}^2
    =
    \frac{
        \nu\widehat{\sigma}_{z,q}^{\,2}
    }{
        \chi^2_{\nu,\epsilon_{\sigma,q}}
    },
    \label{eq:methods_dp_noise_bound}
\end{equation}
where $\chi^2_{\nu,\epsilon_{\sigma,q}}$ denotes the lower-tail
$\epsilon_{\sigma,q}$ quantile of the chi-squared distribution
with $\nu$ degrees of freedom. The statistical derivation of the joint coupling-confidence region and the residual-noise upper bound is provided in Supplementary Information, Section~S1 B.

For each polarization, the parameter-estimation failure budget
is allocated equally among the coupling-region and noise-bound
statements for the two quadratures:
\begin{equation}
    \epsilon_{t,x}=\epsilon_{\sigma,x}
    =\epsilon_{t,p}=\epsilon_{\sigma,p}
    =\frac{\epsilon_{\mathrm{PE}}}{4}.
\end{equation}
By the union bound, all four confidence statements hold
simultaneously with probability at least
$1-\epsilon_{\mathrm{PE}}$.

Let $\boldsymbol{\theta}_H$ collect the
coupling and excess noise parameters for both quadratures, and
let $\mathcal{C}_H$ denote their confidence region. Each candidate
defines a covariance matrix
$\gamma_{A_H A_V B_H}=\gamma_H(\boldsymbol{\theta}_H)$. The worst-case Holevo
information is defined by
\begin{equation}
    \chi_{H,\mathrm{wc}}
    =
    \sup_{\substack{
        \boldsymbol{\theta}_H\in\mathcal{C}_H\\
        \gamma_H(\boldsymbol{\theta}_H)+i\Omega_3\succeq0
    }}
    \chi_H\!\left[
        \gamma_H(\boldsymbol{\theta}_H)
    \right],
    \label{eq:methods_dp_worst_case_holevo}
\end{equation}
where
$\Omega_3=\bigoplus_{k=1}^{3}
\begin{pmatrix}0&1\\-1&0\end{pmatrix}$.
The physicality constraint excludes covariance matrices
incompatible with the uncertainty relation.

In the implemented numerical search, the excess noises of both
quadratures are fixed at their upper confidence bounds, and the
four coupling coefficients are varied jointly over
$\mathcal{E}_x\times\mathcal{E}_p$.

The finite-size key-rate estimate per retained key-generation symbol is
\begin{equation}
    R_H^{\mathrm{finite}}
    =
    \beta_H I(a_H:b_H)
    -
    \chi_{H,\mathrm{wc}}
    -
    \Delta_{N_{\mathrm{rec}}},
    \label{eq:methods_dp_finite_rate}
\end{equation}
where $N_{\mathrm{rec}}$ denotes the number of reconciled symbols
retained for key generation. The observed mutual information is
held fixed during the optimization of the Holevo information.

Statistical uncertainty in parameter estimation is incorporated
into $\chi_{H,\mathrm{wc}}$ through the worst-case covariance
matrix. The additional correction $\Delta_{N_{\mathrm{rec}}}$
accounts for the finite-key penalties associated with
smooth-entropy estimation and privacy amplification:
\begin{equation}
    \Delta_{N_{\mathrm{rec}}}
    =
    (2d_Y+3)
    \sqrt{
        \frac{\log_2(2/\bar{\epsilon})}
             {N_{\mathrm{rec}}}
    }
    +
    \frac{2}{N_{\mathrm{rec}}}
    \log_2\!\left(
        \frac{1}{\epsilon_{\mathrm{PA}}}
    \right),
    \label{eq:methods_dp_finite_correction}
\end{equation}
Here, $\bar{\epsilon}$ is the smoothing parameter,
$\epsilon_{\mathrm{PA}}$ is the privacy-amplification failure
parameter, and $d_Y$ is the dimension parameter associated
with the discretized raw-key variable in the adopted
finite-key bound.

For the simulations in the main text, the parameters are
$V_{M,H}=V_{M,V}=4~\mathrm{SNU}$,
$\varepsilon_H=\varepsilon_V=0.02~\mathrm{SNU}$,
$\beta_H=\beta_V=0.956$,
$N_{\mathrm{rec}}=10^9$,
$p_{\mathrm{EC}}=0.90$,
$\epsilon_{\mathrm{PE}}=\bar{\epsilon}
=\epsilon_{\mathrm{PA}}=10^{-10}$, and $d_Y=2$.
Both polarization channels have the same transmittance,
$T_H=T_V=10^{-\alpha L/10}$, where $L$ is the transmission distance and
$\alpha=0.2~\mathrm{dB/km}$ is the fiber attenuation coefficient.
Ideal detectors are assumed ($\eta_d=1$ and $v_{\mathrm{el}}=0$).

For the experimental rate estimates, the parameters are
$\beta_H=\beta_V=0.96$,
$N_{\mathrm{rec}}=2.5\times10^7$,
$\epsilon_{\mathrm{PE}}=\bar{\epsilon}
=\epsilon_{\mathrm{PA}}=10^{-10}$, and $d_Y=2$. 
A trusted-detector model is used with calibrated detection efficiencies of
$\eta_{d,H}=\eta_{d,V}=0.334$.

The same procedure is applied to the V-polarization data.
Negative rate estimates are set to zero and the total rate estimate is
\begin{equation}
    R_{\mathrm{tot}}^{\mathrm{finite}}
    =
    \max\!\left\{0,R_H^{\mathrm{finite}}\right\}
    +
    \max\!\left\{0,R_V^{\mathrm{finite}}\right\}.
    \label{eq:methods_dp_total_rate}
\end{equation}

\begin{backmatter}

\bmsection{Acknowledgments} 
We thank Simone Conton for assistance with polarization measurements over the deployed fibre channel, and with setting up the transmitter bias control. We also thank Yoann Pietri, Eleni Diamanti and Amine Rhouni for helpful discussions about the experimental setup, and Simone Cammarata for discussions during receiver packaging and characterization. In addition, we acknowledge Joris Van Kerrebrouck for his discussions and support during the packaging of both the transmitter and receiver.

\bmsection{Funding} 
The authors acknowledge funding from the European Union’s Horizon Europe research and innovation programme through the Quantum Security Networks Partnership (QSNP, Grant Agreement No. 101114043) and from Innovation Fund Denmark through the CyberQ project (Grant Agreement No. 3200-00035B). A.A.E.H., S.H., T.G., H.Q.N. and U.L.A. acknowledge support from the Danish National Research Foundation through the Center for Macroscopic Quantum States (bigQ, DNRF142) and from Innovation Fund Denmark through the AccessQKD project (Grant Agreement No. 43560-00004B). S.H. and T.G. acknowledge support from the Carlsberg Foundation (Grant No. CF21-0466). X.Y. acknowledges support from the FWO-Weave SQOPE project (Grant No. G092922N).

\bmsection{Data availability} Data underlying the results presented in this paper are available from the authors upon reasonable request.

\smallskip

\bmsection{Disclosures} The authors declare no conflicts of interest.

\bmsection{Author contributions statement}
Y.B., S.B., S.H., H.Q.N., and A. S. contributed equally to this work. Y.B. developed the security analysis and key-rate calculations under the supervision of A.A.E.H. and U.L.A. S.H. developed the digital signal processing under the supervision of A.A.E.H. and T.G. H.Q.N. and A.A.E.H. performed the QKD experiments. S.B. designed and characterized the receiver electronics under the supervision of X.Y. A.S. designed and assembled the transmitter, characterized the transmitter, and contributed to the experimental setup and data analysis. C.B. designed the transmitter and receiver photonic integrated circuits, co-developed the photonic-electronic receiver concept with X.Y., and contributed to the design of the receiver electronics. G.J.W., A.A.H.A., J.J., A. S. and A.B. contributed to device assembly, experimental setup and fibre-channel characterization. A.A.E.H. conceived the experiment. Y. B. and A.A.E.H. wrote the manuscript with input from  S.B., S.H., H.Q.N., and A. S.  A.A.E.H., X.Y., U.L.A. and T.G. jointly supervised and coordinated the project. All authors contributed to discussions and interpretation of the results.


\bigskip

\end{backmatter}

\bibliography{sample}

\bibliographyfullrefs{sample}


\clearpage

\twocolumn[
  \begin{center}
    {\LARGE\bfseries
    Supplemental Materials: Multimode continuous-variable
    quantum key distribution with CMOS-compatible
    photonic--electronic integration
    \par}

    \vspace{1em}

    {\normalsize
    Yiming Bian\textsuperscript{1,*},
    Sarah Bastiaens\textsuperscript{2,*},
    Saeed Hakimi\textsuperscript{1,*},
    Huy Quang Nguyen\textsuperscript{1,*},
    Aboobackkar Sidhique\textsuperscript{2,*},
    C\'{e}dric Bruynsteen\textsuperscript{2},
    Govind J. Wanjalkar\textsuperscript{2},
    Ali Al Hadi Amhaz\textsuperscript{2},
    Jasper Jans\textsuperscript{2},
    Axl Bomhals\textsuperscript{2},
    Ulrik L. Andersen\textsuperscript{1},
    Tobias Gehring\textsuperscript{1},
    Xin Yin\textsuperscript{2,\(\ddagger\)},
    and Adnan A.E. Hajomer\textsuperscript{1,\(\dagger\)}
    \par}

    \vspace{0.7em}

    {\small\itshape
    \textsuperscript{1}Center for Macroscopic Quantum States
    (bigQ), Department of Physics, Technical University of
    Denmark, 2800 Kongens Lyngby, Denmark\par
    \textsuperscript{2}Ghent University-imec, IDLab,
    Dep. INTEC, 9052 Ghent, Belgium\par}

    \vspace{0.5em}

    {\small
    \textsuperscript{*}These authors contributed equally
    as first authors.\par
    Corresponding authors:
    \textsuperscript{\(\dagger\)}aaeha@dtu.dk;
    \textsuperscript{\(\ddagger\)}xin.yin@ugent.be
    \par}
  \end{center}
  \vspace{1em}
]

\setcounter{page}{1}
\setcounter{section}{0}
\setcounter{figure}{0}
\setcounter{table}{0}
\setcounter{equation}{0}

\renewcommand{\thepage}{S\arabic{page}}
\renewcommand{\thesection}{S\arabic{section}}
\renewcommand{\thefigure}{S\arabic{figure}}
\renewcommand{\thetable}{S\arabic{table}}
\renewcommand{\theequation}{S\arabic{equation}}

\vspace{1.5em}

\section{Theory}
\subsection{Dual polarization system: Residual crosstalk model}
\begin{figure*}[b]
    \centering
    \includegraphics[width=0.7\linewidth]{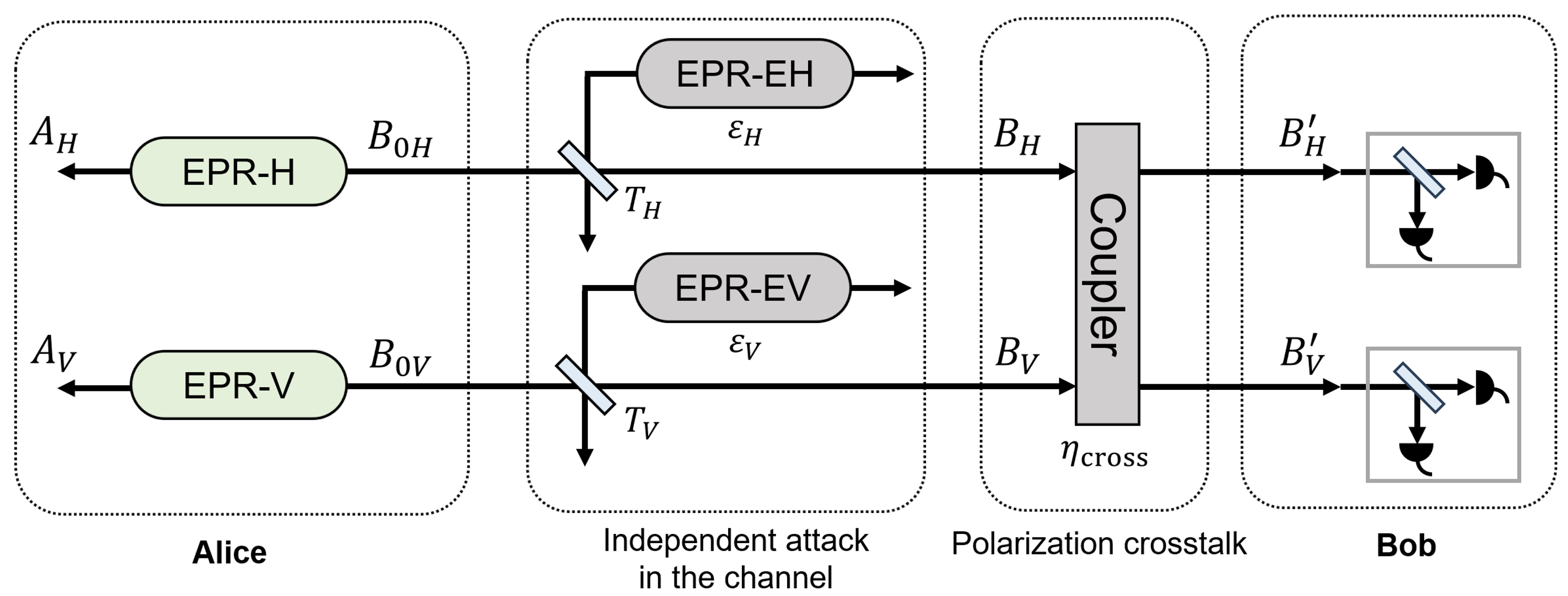}
    \caption{\textbf{Schematic diagram of the dual-polarization CV-QKD model considering polarization crosstalk.} 
    An ideal dual-polarization system functions as two independent single-polarization CV-QKD setups. To account for polarization crosstalk, the transmitted optical modes $B_H$ and $B_V$ are coupled via a beam splitter (Coupler) after experiencing independent channel attacks. Finally, the output modes $B'_H$ and $B'_V$ are measured independently by Bob using two heterodyne detectors. }
    \label{fig:Dual_Pol_Model}
\end{figure*}

We model the polarization crosstalk in a dual-polarization system according to the scheme shown in Fig. \ref{fig:Dual_Pol_Model}. We have two independent EPR sources with output modes suffering from independent attacks on the channel (here we consider independent attacks for simplification). After this, the two output modes, $B_H$ and $B_V$, are combined by a coupler with transmittance $\eta_{\mathrm{cross}}$, which characterizes the polarization leakage in the channel. Finally, the receiver modes, $B_H'$ and $B_V'$, are heterodyne detected, respectively. 
Here, $\eta_{\mathrm{cross}}$ denotes the fraction of optical power retained in the intended polarization after polarization compensation, whereas $1-\eta_{\mathrm{cross}}$ quantifies the residual power leakage from the orthogonal polarization. In the ideal case of perfect polarization recovery, $\eta_{\mathrm{cross}}=1$. In practical implementations, however, imperfect polarization recovery generally results in $\eta_{\mathrm{cross}}<1$.

Focusing on Pol-H, we have
\begin{equation}
  B_H' = \sqrt{\eta_{\mathrm{cross}}}\,B_H
  + \sqrt{1-\eta_{\mathrm{cross}}}\,B_V.
\end{equation}
Therefore,
\begin{equation}
\begin{aligned}
  \left\langle {B_H'}^2 \right\rangle
  ={}& \eta_{\mathrm{cross}}\left\langle B_H^2 \right\rangle
  + \left(1-\eta_{\mathrm{cross}}\right)\left\langle B_V^2 \right\rangle \\
  &+ 2\sqrt{\eta_{\mathrm{cross}}\left(1-\eta_{\mathrm{cross}}\right)}
  \left\langle B_H,B_V \right\rangle.
\end{aligned}
\end{equation}
The independent attack indicates that $\left\langle B_H,B_V \right\rangle=0$; therefore,
\begin{equation}
  \left\langle {B_H'}^2 \right\rangle
  = \eta_{\mathrm{cross}}\left\langle B_H^2 \right\rangle
  + \left(1-\eta_{\mathrm{cross}}\right)\left\langle B_V^2 \right\rangle.
\end{equation}
Denote the modulation variance in Pol-H and Pol-V as $V_{MH}$ and $V_{MV}$, and suppose that
\begin{equation}
  \left\langle B_H^2 \right\rangle
  = T_H\left(V_{MH}+\varepsilon_H\right)+1,
  \quad
  \left\langle B_V^2 \right\rangle
  = T_V\left(V_{MV}+\varepsilon_V\right)+1.
\end{equation}
We then have
\begin{equation}
\begin{aligned}
\left\langle {B_H'}^2 \right\rangle
={}& \eta_{\mathrm{cross}}
\left[T_H\left(V_{MH}+\varepsilon_H\right)+1\right] \\
&+\left(1-\eta_{\mathrm{cross}}\right)
\left[T_V\left(V_{MV}+\varepsilon_V\right)+1\right] \\
={}& \eta_{\mathrm{cross}}T_H
\left(V_{MH}+\varepsilon_H\right)
+\eta_{\mathrm{cross}}
+T_V\left(V_{MV}+\varepsilon_V\right)+1 \\
&-\eta_{\mathrm{cross}}
\left[T_V\left(V_{MV}+\varepsilon_V\right)+1\right] \\
={}& \eta_{\mathrm{cross}}T_H
\left(V_{MH}+\varepsilon_H\right)
+T_V\left(V_{MV}+\varepsilon_V\right) \\
&-\eta_{\mathrm{cross}}T_V
\left(V_{MV}+\varepsilon_V\right)+1 \\
={}& \eta_{\mathrm{cross}}T_H\Biggl[
V_{MH}+\varepsilon_H
-\frac{T_V}{T_H}\left(V_{MV}+\varepsilon_V\right) \\
&\qquad\qquad
+\frac{T_V}{\eta_{\mathrm{cross}}T_H}
\left(V_{MV}+\varepsilon_V\right)
\Biggr]+1.
\end{aligned}
\end{equation}

We define the effective H-polarization transmittance as
$T_{H,\mathrm{eff}}=\eta_{\mathrm{cross}}T_H$ and write the
output variance as
$\langle {B_H'}^2\rangle
=1+T_{H,\mathrm{eff}}(V_{MH}+\varepsilon_{\mathrm{eq}})$.
The equivalent excess noise, referred to the input of this
effective channel, is therefore
\begin{equation}
\begin{aligned}
  \varepsilon_{\mathrm{eq}}
  &=
  \varepsilon_H
  +\frac{T_V}{T_H}
  \frac{1-\eta_{\mathrm{cross}}}{\eta_{\mathrm{cross}}}
  \left(V_{MV}+\varepsilon_V\right),\\
  &\qquad T_H>0,\quad
  0<\eta_{\mathrm{cross}}\leq 1.
\end{aligned}
\end{equation}
The additional contribution contains both the modulation and
the excess noise carried by the orthogonal polarization.

For equal channel transmittances, $T_H=T_V$, this expression
reduces to
\begin{equation}
  \varepsilon_{\mathrm{eq}}
  =
  \varepsilon_H
  +\frac{1-\eta_{\mathrm{cross}}}{\eta_{\mathrm{cross}}}
  \left(V_{MV}+\varepsilon_V\right).
  \label{eq:S7}
\end{equation}
If the two channels also have equal input-referred excess noise,
$\varepsilon_H=\varepsilon_V=\varepsilon$, we obtain
\begin{equation}
  \varepsilon_{\mathrm{eq}}
  =
  \frac{1-\eta_{\mathrm{cross}}}{\eta_{\mathrm{cross}}}V_{MV}
  +\frac{\varepsilon}{\eta_{\mathrm{cross}}}.
\end{equation}

For single-polarization encoding, setting $V_{MV}=0$ removes
the contribution from modulation in the orthogonal polarization.
Under the equal-noise assumption above, the unmodulated V mode
still carries channel excess noise, giving
$\varepsilon_{\mathrm{eq}}=\varepsilon/\eta_{\mathrm{cross}}$.
Here, $1/\eta_{\mathrm{cross}}$ is an input-referencing factor,
while the effective H-polarization transmittance is
$\eta_{\mathrm{cross}}T_H$.
For $\eta_{\mathrm{cross}}=0.98$--$0.99$, this factor increases
the equivalent excess noise by only approximately 1--2\%.
If the orthogonal input to the coupler is instead a pure vacuum
mode, with $V_{MV}=\varepsilon_V=0$, the equivalent excess noise
remains $\varepsilon_H$, while the effective transmittance is
reduced.

\begin{figure}
    \centering
    \includegraphics[width=0.9\linewidth]{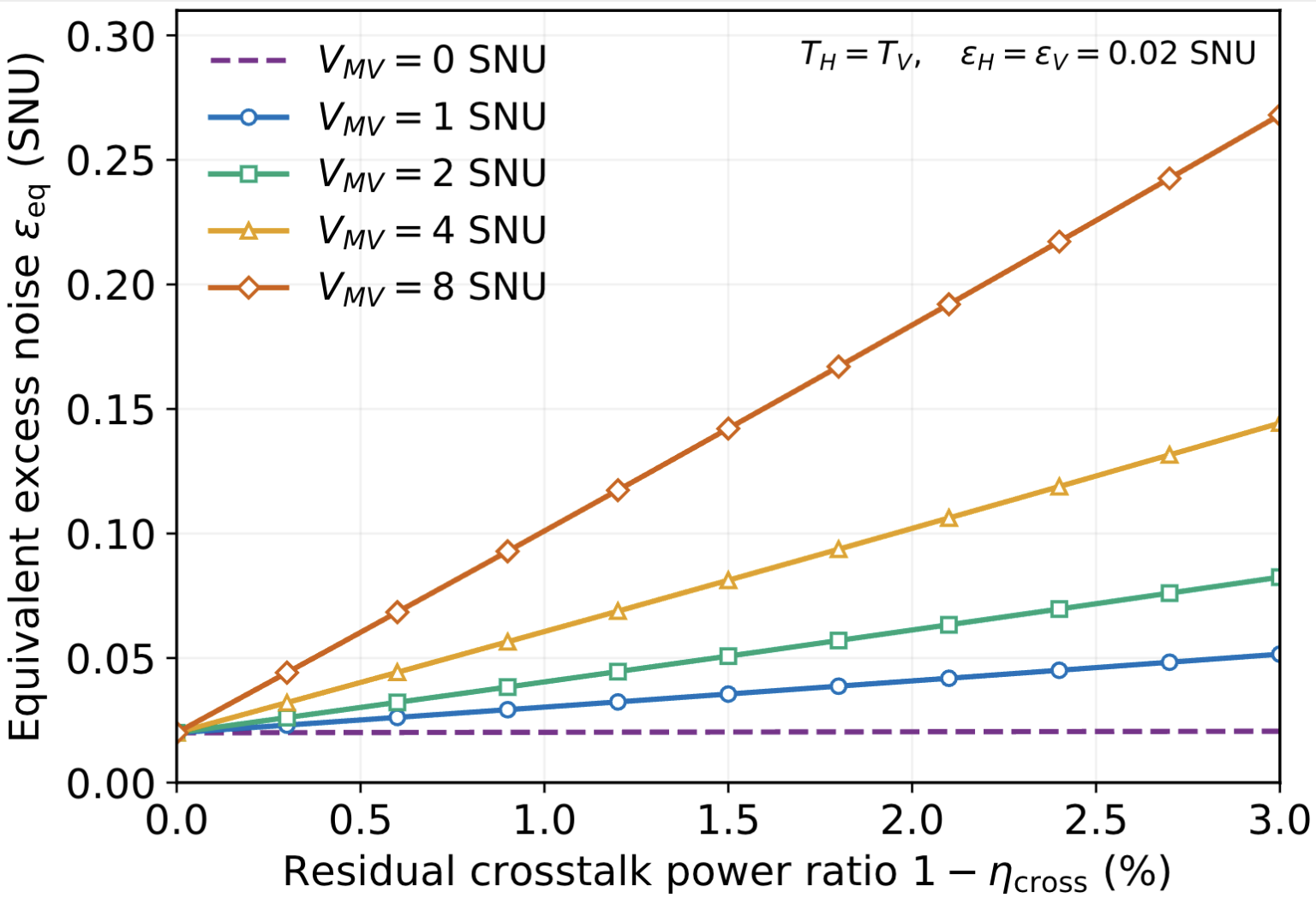}
    \caption{\textbf{Effect of residual polarization crosstalk
    on the equivalent excess noise.}
    Equivalent input-referred excess noise of the H-polarization
    channel, $\varepsilon_{\mathrm{eq}}$, versus residual
    crosstalk power ratio, $1-\eta_{\mathrm{cross}}$, calculated
    using \eqref{eq:S7} for different V-polarization modulation
    variances $V_{MV}$.
    Parameters are $T_H=T_V$ and
    $\varepsilon_H=\varepsilon_V=0.02$~SNU.
    Open symbols distinguish nonzero modulation variances,
    while the dashed curve represents $V_{MV}=0$.}
    \label{fig:Equivalent_Excess_Noise}
\end{figure}

By contrast, when the orthogonal V polarization carries a
modulated signal ($V_{MV}>0$), even a small residual crosstalk
power ratio can produce a substantial increase in the
equivalent excess noise.
Figure~\ref{fig:Equivalent_Excess_Noise} illustrates how residual
polarization crosstalk converts modulation in the orthogonal
polarization into additional equivalent excess noise.
For a fixed nonzero residual crosstalk power ratio and fixed
channel excess noise, the equivalent excess noise increases
linearly with $V_{MV}$, as predicted by \eqref{eq:S7}.
This dependence introduces a practical trade-off in
dual-polarization CV-QKD: increasing the modulation variance
in one polarization can improve its received signal-to-noise
ratio and facilitate error correction, while increasing the
crosstalk-induced noise penalty in the orthogonal polarization.
For example, with $\varepsilon_H=\varepsilon_V=0.02$~SNU,
a residual crosstalk power ratio of only $1\%$ raises the
H-polarization equivalent excess noise to approximately
$0.061$~SNU when $V_{MV}=4$~SNU, about three times its value
without crosstalk.
Modulation variances should therefore be selected jointly
with the residual crosstalk level, accounting for both
reconciliation performance and the noise penalty imposed
on the orthogonal polarization.

\begin{figure}[t]
    \centering
    \includegraphics[width=0.95\linewidth]{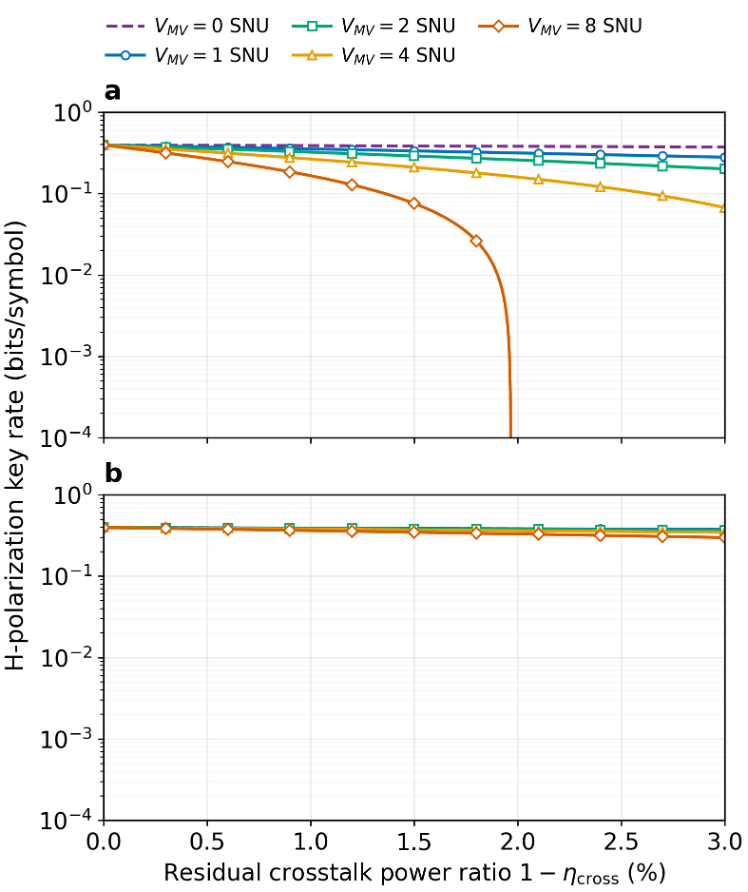}
    \caption{\textbf{Impact of residual polarization crosstalk on the secret key rate under single-mode and multimode security analyses.} \textbf{a,} Single-mode analysis, which treats the contribution from the orthogonal polarization as equivalent excess noise. \textbf{b,} Multimode analysis, which retains the correlations in $\gamma_{A_HA_VB_H}$ while conservatively assigning the unused receiver mode $B_V$ to Eve. Both analyses use the same mutual information $I(a_H:b_H)$. Parameters are $L=10$~km, $T_H=T_V$, $V_{MH}=4$~SNU, $\varepsilon_H=\varepsilon_V=0.02$~SNU, $\beta=0.956$, and ideal heterodyne detection.}
    \label{fig:SKR_Comparison}
\end{figure}

This trade-off also motivates a multimode security framework that explicitly retains correlations arising from inter-mode coupling. In a single-mode description, the contribution from the orthogonal polarization is incorporated into the equivalent excess noise without retaining its correlation with Alice's modulation in that polarization. As shown in Fig.~\ref{fig:SKR_Comparison}\textbf{a}, increasing $V_{MV}$ therefore leads to a more pronounced reduction in the H-polarization secret key rate as residual crosstalk increases. By including both trusted source modes in the security analysis, the multimode framework accounts for these correlations when bounding Eve's information and substantially reduces the key-rate penalty, as illustrated in Fig.~\ref{fig:SKR_Comparison}\textbf{b}. For example, at a residual crosstalk power ratio of $1\%$ and $V_{MV}=4$~SNU, the three-mode analysis increases the estimated rate from approximately $0.266$ to $0.379$~bits/symbol, compared with $0.396$~bits/symbol without crosstalk. Both analyses use the same mutual information $I(a_H:b_H)$, so the improvement arises from a tighter bound on Eve's Holevo information rather than an enhancement of the reconciliation term. This improvement does not require treating the coupling as trusted noise: the channel and its residual mode coupling remain untrusted. The multimode security analysis is detailed below.

The three-mode formulation retains the useful correlations between Alice's two polarization components and Bob's selected key mode, while conservatively assigning Bob's unused polarization mode to Eve. It therefore exploits the multipoint-to-point correlation generated by residual polarization mixing without assuming that the residual crosstalk itself is trusted. Compared with an independent two-mode analysis based only on $\gamma_{A_HB_H}$, the present model preserves more experimentally accessible correlation information. 

\subsection{Statistical derivation of the joint parameter-estimation bounds}
\label{sec:multipoint-to-point-finite-size}

This section derives the joint confidence region for the coupling coefficients and the one-sided upper confidence bound on the residual-noise variance used in Methods. The derivation is conditional on the observed modulation samples and assumes the Gaussian linear regression model specified there.

For each quadrature $q\in\{x,p\}$, write the H-polarization measurement model in vector form as
\begin{equation}
    \mathbf{b}_{H,q} = X_q\mathbf{t}_q+\mathbf{z}_q.
\end{equation}
The measurement vector, design matrix, coupling vector and residual-noise vector are defined by
\begin{equation}
    \underbrace{\begin{pmatrix}
        b_{H,1}^{q}\\
        \vdots\\
        b_{H,N_{\mathrm{rec}}}^{q}
    \end{pmatrix}}_{\mathbf{b}_{H,q}}
    =
    \underbrace{\begin{pmatrix}
        a_{H,1}^{q} & a_{V,1}^{q}\\
        \vdots & \vdots\\
        a_{H,N_{\mathrm{rec}}}^{q} & a_{V,N_{\mathrm{rec}}}^{q}
    \end{pmatrix}}_{X_q}
    \underbrace{\begin{pmatrix}
        t_{HH}^{q}\\
        t_{VH}^{q}
    \end{pmatrix}}_{\mathbf{t}_q}
    +
    \underbrace{\begin{pmatrix}
        z_1^{q}\\
        \vdots\\
        z_{N_{\mathrm{rec}}}^{q}
    \end{pmatrix}}_{\mathbf{z}_q}.
    \label{eq:finite-regression-expanded}
\end{equation}
Here, each row of $X_q$ corresponds to one sample, and its two columns contain Alice's H- and V-polarization modulation samples, respectively. For a block of $N_{\mathrm{rec}}$ samples, the matrix $X_q\in\mathbb{R}^{N_{\mathrm{rec}}\times2}$ is assumed to have full column rank. 

The residual-noise vector $\mathbf{z}_q$ is assumed to be independent of Alice's modulation data $X_q$ and to consist of independent, identically distributed zero-mean Gaussian samples with variance $\sigma_{z,q}^{2}$. Consequently,
\begin{equation}
    \mathbf{z}_q\mid X_q \sim \mathcal{N}\!\left(\mathbf{0},\sigma_{z,q}^{2}\mathbf{I}_{N_{\mathrm{rec}}}\right).
\end{equation}
Here, $\sigma_{z,q}^{2}$ denotes the unknown residual-noise variance per sample for quadrature $q$, and $\mathbf{I}_{N_{\mathrm{rec}}}$ is the $N_{\mathrm{rec}}\times N_{\mathrm{rec}}$ identity matrix. Since the noise is independent of the modulation data, conditioning on $X_q$ leaves its distribution unchanged. 

Substituting the above model into the least-squares estimator in Eq.~(\ref{eq:methods_dp_estimators}) gives
\begin{equation}
    \widehat{\mathbf{t}}_q-\mathbf{t}_q = \left(X_q^{\mathsf T}X_q\right)^{-1}X_q^{\mathsf T}\mathbf{z}_q.
\end{equation}
Consequently,
\begin{equation}
    \widehat{\mathbf{t}}_q-\mathbf{t}_q \sim \mathcal{N}_2\!\left(\mathbf{0},\sigma_{z,q}^{2}\left(X_q^{\mathsf T}X_q\right)^{-1}\right).
    \label{eq:finite-ols-error-distribution}
\end{equation}
The two coupling estimates are therefore treated jointly, with their covariance determined by $\left(X_q^{\mathsf T}X_q\right)^{-1}$. The standardized quadratic estimation error satisfies
\begin{equation}
    Q_{t,q} \equiv \frac{\left(\mathbf{t}_q-\widehat{\mathbf{t}}_q\right)^{\mathsf T}X_q^{\mathsf T}X_q\left(\mathbf{t}_q-\widehat{\mathbf{t}}_q\right)}{\sigma_{z,q}^{2}} \sim\chi_2^2,
    \label{eq:finite-coupling-chi-square}
\end{equation}
where the two degrees of freedom correspond to the two jointly estimated coupling coefficients.

To characterize the residual variance, define the orthogonal projection onto the column space of $X_q$ as
\begin{equation}
    P_q = X_q\left(X_q^{\mathsf T}X_q\right)^{-1}X_q^{\mathsf T}.
\end{equation}
The fitted residual vector is
\begin{equation}
    \mathbf{r}_q = \mathbf{b}_{H,q}-X_q\widehat{\mathbf{t}}_q = \left(\mathbf{I}_{N_{\mathrm{rec}}}-P_q\right)\mathbf{z}_q.
\end{equation}
Since $P_q$ has rank two, the residual projection has rank $\nu=N_{\mathrm{rec}}-2$. The normalized residual sum of squares therefore satisfies
\begin{equation}
    Q_{\sigma,q} \equiv \frac{\|\mathbf{r}_q\|_2^2}{\sigma_{z,q}^{2}} = \frac{\nu\widehat{\sigma}_{z,q}^{\,2}}{\sigma_{z,q}^{2}} \sim\chi_\nu^2.
    \label{eq:finite-residual-chi-square}
\end{equation}
This also explains the denominator $N_{\mathrm{rec}}-2$ in the unbiased residual-variance estimator used in Methods.

Under the Gaussian-noise assumption, the fitted and residual components are independent because they arise from orthogonal projections of $\mathbf{z}_q$. Thus, $Q_{t,q}$ and $Q_{\sigma,q}$ are independent, and
\begin{equation}
    \frac{Q_{t,q}/2}{Q_{\sigma,q}/\nu} = \frac{\left(\mathbf{t}_q-\widehat{\mathbf{t}}_q\right)^{\mathsf T}X_q^{\mathsf T}X_q\left(\mathbf{t}_q-\widehat{\mathbf{t}}_q\right)}{2\widehat{\sigma}_{z,q}^{\,2}} \sim F_{2,\nu}.
    \label{eq:finite-regression-f-statistic}
\end{equation}
The unknown variance $\sigma_{z,q}^{2}$ cancels in this ratio. Taking the $(1-\epsilon_{t,q})$ quantile of $F_{2,\nu}$ therefore gives the joint confidence ellipsoid $\mathcal{E}_q$ in Eq.~(\ref{eq:methods_dp_confidence_region}), with
\begin{equation}
    \Pr\!\left[\mathbf{t}_q\in\mathcal{E}_q\mid X_q\right] = 1-\epsilon_{t,q}.
\end{equation}
This coverage statement is exact under the stated Gaussian regression model.

The one-sided upper confidence bound on the residual variance follows from Eq.~(\ref{eq:finite-residual-chi-square}). Let $\chi^2_{\nu,\epsilon_{\sigma,q}}$ denote the lower-tail $\epsilon_{\sigma,q}$ quantile, such that
\begin{equation}
    \Pr\!\left[Q_{\sigma,q}\leq\chi^2_{\nu,\epsilon_{\sigma,q}}\mid X_q\right] = \epsilon_{\sigma,q}.
\end{equation}
Rearranging the complementary event yields
\begin{equation}
    \Pr\!\left[\sigma_{z,q}^{2}\leq\frac{\nu\widehat{\sigma}_{z,q}^{\,2}}{\chi^2_{\nu,\epsilon_{\sigma,q}}}\,\middle|\,X_q\right] = 1-\epsilon_{\sigma,q},
\end{equation}
which establishes the upper confidence bound in Eq.~(\ref{eq:methods_dp_noise_bound}).

\begin{figure*}[t]
    \centering
    \includegraphics[width=1\linewidth]{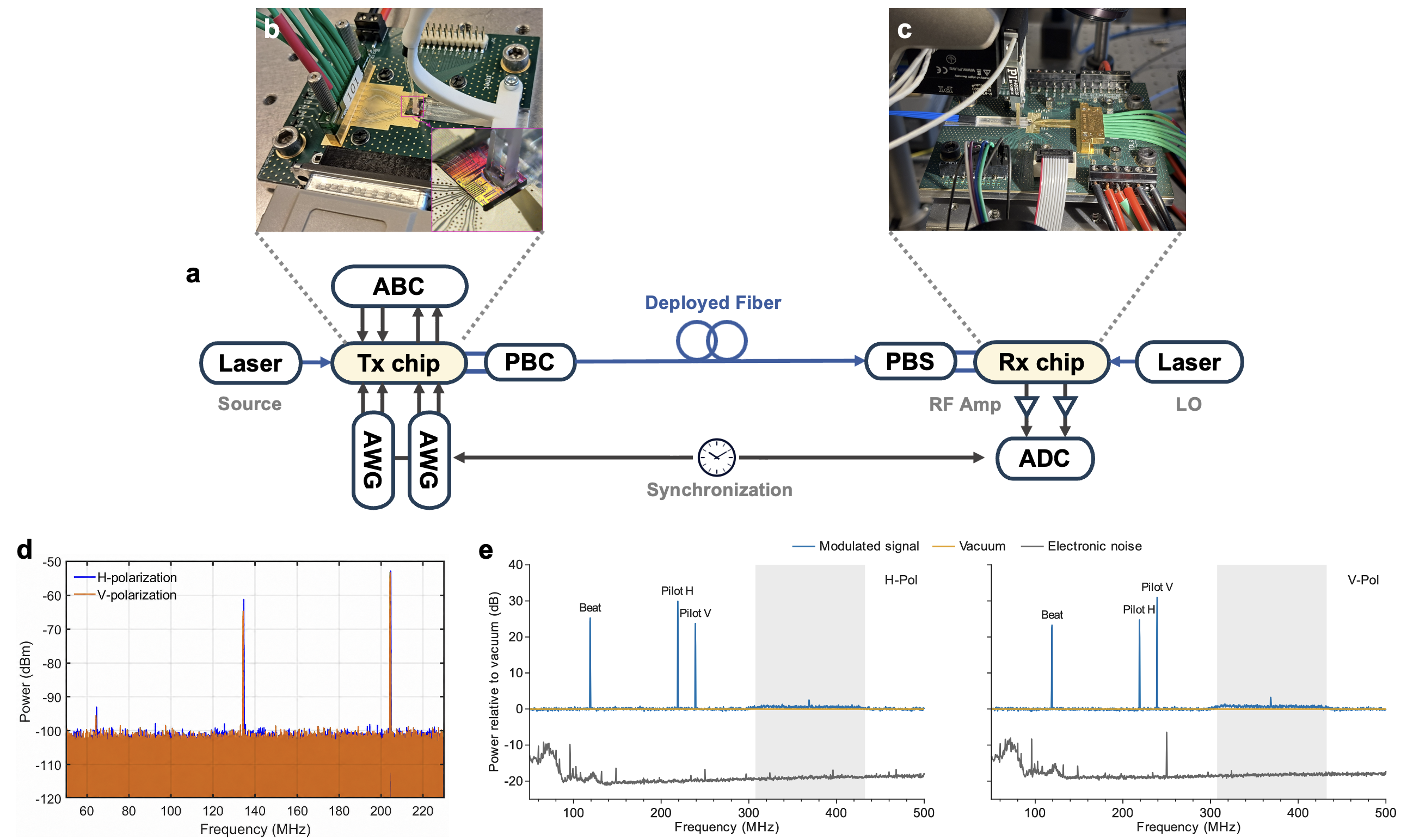}
    \caption{Experimental configuration and packaged chip-based devices. \textbf{a,} System schematic. \textbf{b,} Photograph of the packaged transmitter; the inset shows a close-up of the chip and fiber-array interface. \textbf{c,} Photograph of the receiver assembly, comprising the photonic chip, transimpedance amplifier and fiber-array coupling arrangement. \textbf{d,} Measured spectra used to characterize sideband suppression in the H- and V-polarization IQ modulators. \textbf{e,} Measured spectra of the H- and V-polarization channels, normalized to vacuum noise, showing the beat and pilot tones and the quantum-signal bands (shaded). Tx/Rx, transmitter/receiver; AWG, arbitrary waveform generator; ABC, automatic bias controller; IQ, in-phase and quadrature; PBC/PBS, polarization beam combiner/splitter; LO, local oscillator; ADC, analog-to-digital converter. Blue lines indicate optical paths, and dark-grey lines indicate electrical signal, control and synchronization connections.}
    \label{fig:System}
\end{figure*}

\section{System}

Figure~\ref{fig:System}a summarizes the optical and electrical connections of the experimental system, with photographs of the packaged transmitter and receiver shown in Figs.~\ref{fig:System}b and c, respectively. At the transmitter, two arbitrary waveform generators (AWGs) drive the IQ modulators, while an automatic bias controller (ABC) maintains their operating points. The two optical outputs are polarization-multiplexed for fiber transmission and separated at the receiver for parallel detection. The amplified electrical outputs are digitized for offline processing.

The transmitter package in Fig.~\ref{fig:System}b uses a custom printed circuit board (PCB) with four differential modulation pairs and DC connections for bias and attenuation control. The differential pairs have approximately equal routing lengths to reduce timing skew. On-chip waveguide loop-back structures facilitate fiber-array alignment. After alignment, the fiber array is fixed with adhesive. The measured insertion loss through the outer loop-back path is approximately 6~dB.

Figure~\ref{fig:System}d presents the sideband-suppression measurements. Each IQ modulator is driven by 70-MHz sinusoidal I/Q signals, and its optical output is combined with an independent local oscillator using an optical coupler. The resulting beat signal is detected by a high-speed photodiode and recorded with an electrical spectrum analyzer. The laser-frequency offset is selected to resolve both modulation sidebands without spectral folding. The measured sideband suppression exceeds 30~dB for both modulators.

In the receiver assembly shown in Fig.~\ref{fig:System}c, the photonic and electronic dies are mounted on a common custom PCB. The noise measurements include the complete electrical readout chain, comprising the TIA, external RF amplifier and ADC.

Figure~\ref{fig:System}e compares the modulated-signal, vacuum-noise and electronic-noise spectra in the two receiver channels, normalized to the corresponding vacuum-noise levels. The vacuum-noise reference is acquired with the LO present and the signal input blocked; the electronic-noise reference is acquired without optical input. The shaded regions indicate the quantum-signal bands.

\ifthenelse{\equal{\journalref}{aop}}{%
\section*{Author Biographies}
\begingroup
\setlength\intextsep{0pt}
\begin{minipage}[t][6.3cm][t]{1.0\textwidth} 
  \begin{wrapfigure}{L}{0.25\textwidth}
    \includegraphics[width=0.25\textwidth]{john_smith.eps}
  \end{wrapfigure}
  \noindent
  {\bfseries John Smith} received his BSc (Mathematics) in 2000 from The University of Maryland. His research interests include lasers and optics.
\end{minipage}
\begin{minipage}{1.0\textwidth}
  \begin{wrapfigure}{L}{0.25\textwidth}
    \includegraphics[width=0.25\textwidth]{alice_smith.eps}
  \end{wrapfigure}
  \noindent
  {\bfseries Alice Smith} also received her BSc (Mathematics) in 2000 from The University of Maryland. Her research interests also include lasers and optics.
\end{minipage}
\endgroup
}{}

\end{document}